# Broadband Solar Selective Absorber with Dallenbach-type Bilayer Structure Achieved Using Carbon Nanotube Membranes with Tailored Optical Spectra

*Hengkai Wu[1], Taishi Nishihara[1,2,3], Mioko Hizukuri[1], Takeshi Tanaka[4], Hiromichi Kataura[5], Yuhei Miyauchi[1,6,\*]*

[1] Institute of Advanced Energy, Kyoto University, Uji, Kyoto 611-0011, Japan

[2] Department of Physics, Faculty of Science, Tokyo University of Science, Shinjuku, Tokyo 162-8601, Japan

[3] Research Institute for Science and Technology, Tokyo University of Science, Shinjuku, Tokyo 162-8601, Japan

[4] Research Institute of Core Technology for Materials Innovation, National Institute of Advanced Industrial Science and Technology (AIST), Tsukuba, Ibaraki 305-8565, Japan

[5] Nanomaterials Research Institute, National Institute of Advanced Industrial Science and Technology (AIST), Tsukuba, Ibaraki 305-8565, Japan

[6] Department of Mechanical Engineering, The University of Tokyo, 7-3-1 Hongo, Bunkyo-ku, Tokyo 113-8656, Japan

*Correspondence to: ymiyauchi@g.ecc.u-tokyo.ac.jp

ABSTRACT: Spectrally selective absorbers that maximize solar absorption and minimize thermal radiation loss are crucial for efficient solar thermal energy harvesting. However, limitations imposed by the intrinsic properties of conventional materials hinder the fabrication of interference-based absorbers with desired optical properties. Herein, a high-performance solar absorber with a Dallenbach-type dielectric–metal tandem structure was fabricated using an ultrathin, subquarter-wavelength thickness single-walled carbon nanotube (SWCNT) membrane with tailored optical spectra as the absorbing layer. By mixing multiple SWCNT chiral structures, the optical response of the absorbing layer was tailored to approximate a theoretical complex refractive index spectrum of dielectrics; therefore enabling high solar absorptance, low infrared emittance, and a low angular dependence with a simple bilayer structure. The fabricated proof-of-concept 1-cm$^2$ absorber exhibited excellent spectral selectivity (solar absorptance/infrared emittance of 0.84/0.03), and an equilibrium temperature of ≈190 ºC (270 ºC) under nonconcentrated (2× concentrated) sunlight, considerably outperformed a blackbody-like absorber. This study presents a high-performance solar absorber with a simple planar structure and proposes a concept of designing dielectric with the desired optical properties by mixing various types of structure-sorted SWCNTs.

Efficient solar energy harvesting is crucial for addressing global energy challenges and realizing a carbon-neutral society. Considering solar thermal energy harvesting utilizes broadband solar energy with high efficiency and has potential in implementation costs, it has garnered sustained attention. During this process, sunlight is converted into thermal energy for many applications such as power generation, space heating and cooling, and seawater desalination.[1–6] Spectrally selective absorbers (SSAs) have been developed to efficiently extract thermal energy from sunlight without relying on large-scale, costly light-concentration equipment.[7–14] Figure 1 presents the fundamental design principle of a solar absorber. As shown in Figure 1a, despite a blackbody-like absorber can completely absorbs sunlight, it also emits thermal radiation in the infrared region efficiently based on Kirchhoff's law of radiation—spectral absorptance ($\alpha_\lambda$) is equal to emittance ($\varepsilon_\lambda$) for a body in thermal equilibrium. The strong thermal radiation decreases the equilibrium temperature and reduces both the quality and amount of usable thermal energy. An ideal SSA (Figure 1b) minimizes the thermal radiation losses while retaining the high solar absorptance, thereby facilitating thermal energy extraction. Such SSAs require near-unity absorptance across the solar spectrum (mainly 0.3–2.5 μm), low absorptance (and thus low emittance) in the infrared region, and an appropriate absorption cutoff in the near-to-shortwave infrared region where absorptance rapidly transitions from one to zero, with low sensitivity to the angle of incidence. The details about cutoff wavelength can be found in Supplementary Note 1, Figures. S1 and S2.

Because a single material cannot perfectly meet all the aforementioned requirements, various composite structures have been introduced to modify the absorption characteristics, including absorber–reflector tandems, multilayers, textured surface, cavity structures, and photonic crystals.[1,4,12–25] Among them, the absorber–reflector tandem (upper panel of Figure 1c) is the simplest composite structure for an SSA. The absorber layer, typically thicker than the wavelength of the light to be absorbed, contribute to the main absorption feature, and the reflector layer enhances absorption by doubling the optical path length. Absorptive materials such as transition metals and their nitrides, oxides, and borides have been used as the absorber layer for broadband high absorptance.[26–34] Although these materials exhibit high sunlight absorption, they also show unsharp cutoff and non-negligible infrared emittance, leading to unsatisfactory spectral selectivity. As an alternative, semiconductors have also been used as absorber layers because their bandgaps provide a sharp absorption cutoff and enable low infrared emittance.[35–38] However, semiconductors have disadvantages inherent to their physical properties: their high refractive index *n* causes high reflection losses and insufficient solar absorption, necessitating the incorporation of one or more antireflection layers. This complicates the manufacturing process and lead to an unwanted increase in the angular sensitivity of absorption features. In addition, the thermally generated carriers at high temperatures enhance the infrared absorptance and emittance,[39–41] thereby degrading the performance of semiconductor-based absorbers.

Herein, we propose a single-walled carbon nanotube (SWCNT)-based high-performance SSA with a simple absorber–reflector tandem structure. The key concept in the design strategy was to place an ultrathin, subquarter-wavelength semiconducting layer with tailored absorption characteristics on a metal reflector (bottom panel of Figure 1c). This structure is conventionally known as a Dallenbach absorber,[42–45] in which a lossy thin film simultaneously functions as an absorber and an antireflection dielectric. Conventional Dallenbach absorbers are usually regarded as resonant absorbers, exhibiting high absorptance only at certain wavelengths. We addressed this limitation by tailoring the optical response of the absorbing layer. We demonstrated that an ultrathin absorbing layer with the required optical characteristics can be realized by mixing various semiconducting SWCNT structures with different optical spectra and that high-performance solar absorbers with a nearly ideal broadband optical response and low angular sensitivity can be designed and fabricated. As a proof-of-concept, a solar absorber comprising a ~100-nm thick mixed-SWCNT absorbing layer, with the tailored optical spectrum, deposited on a gold-coated substrate with an area of 1 cm$^2$ was fabricated. This absorber exhibited outstanding spectral selectivity (absorptance/emittance of 0.84/0.03) and equilibrium temperature of ≈ 190 ºC (270 ºC) under nonconcentrated (2× concentrated) sunlight, considerably outperforms that of a blackbody-like absorber by ≈ 100 ºC (150 ºC).

## RESULT AND DISCUSSION

**Concept of absorber design.** Figure 2a shows the fundamental design concept of the proposed absorber structure, which is based on the concept of the Dallenbach absorber. The design concept can be elucidated by considering the optical interference at a specific wavelength $\lambda$ through a material with complex refractive index $\tilde{n}(\lambda) = n(\lambda) + i\kappa(\lambda)$, where $n$ and $\kappa$ are the real and imaginary parts, respectively. The membrane with a thickness $d = \lambda/4n$ gives rise to destructive interference of the reflections at the air–membrane interface and the membrane–metal interface, similar to optical antireflection coatings.[46] However, the key difference is that the layer comprises a lossy dielectric. Considering the complex amplitude reflection coefficient $r_1 = (1 - \tilde{n})/(1 + \tilde{n})$ at the first interface, the reflected electric field is reduced by $|r_1|$ times. Meanwhile, the light electric field component propagating through the absorbing layer, reflected by the metal reflecting surface (here we consider the metal as a perfect conductor for simplicity), and returning to the first interface is absorbed by a factor of $\exp(-4\pi\kappa d/\lambda) = \exp(-\pi\kappa/n)$ after the round trip of $2d$. This round trip introduces a phase shift of approximately $\pi$ to the reflected field, causing the destructive interference with the reflected field from the first interface. Therefore, if a semiconductor has $n$ and $\kappa$ satisfying $|r_1| \approx \exp(-\pi\kappa/n)$ at $\lambda$, near-perfect absorption can be achieved at this wavelength. The required $\kappa$ can be analytically obtained as $\kappa \approx 2/\pi$, regardless of $\lambda$ (see Supplementary Note 2 and Figure S3 for the derivation).[43,44] Regarding the real part $n$, because the membrane thickness $d$ is a constant, $n$ should be proportional to the wavelength as $n \approx \lambda/4d$. Therefore, if a material fulfills $\kappa \approx 2/\pi$ and $n \approx \lambda/4d$ for $\lambda \leq \lambda_{\text{cutoff}}$ and $\kappa \approx 0$ for $\lambda > \lambda_{\text{cutoff}}$, the absorber can be a nearly ideal SSA with a broad absorptance spectrum and a sharp cutoff at $\lambda_{\text{cutoff}}$. Figure 2b shows the simulated reflectance ($R$) and absorptance ($A = 1 - R$) spectra of a Dallenbach absorber with a hypothetical absorbing dielectric ($d = 110$ nm) with $n$ and $\kappa$ that approximately fulfills the aforementioned requirements. Herein, the modeled $n$ and $\kappa$ shown in the inset was constructed to satisfy the Kramers-Kronig relation by considering the summation of the responses of multiple Lorentz oscillators (Supplementary Note 3 and Figure S4). The designed $n$ and $\kappa$ approximately fulfill $n \approx \lambda/4d$ and $\kappa \approx 2/\pi$ for $\lambda \leq \lambda_{\text{cutoff}}$ and $\kappa \approx 0$ for $\lambda > \lambda_{\text{cutoff}}$, respectively, as indicated by the blue- and red-filled regions in the inset; thus, near-unity absorptance in a broadband region over visible-to-near infrared region and a sharp cut off at $\lambda_{\text{cutoff}}$ are achieved.

**Design of carbon nanotube-based selective absorber.** The design principle for the proposed structure is simple; however, achieving $n \approx \lambda/4d$ and a constant $\kappa \approx 2/\pi$ in the broadband solar spectral range using conventional semiconductors is difficult. Figure 3a shows the complex refractive index spectra of some typical semiconductors.[47–49] For these materials, $\kappa \approx 2/\pi$ can be achieved only at a specific wavelength, and $\kappa$ decreases with increasing wavelength; $n$ is also larger than the desired value. However, semiconducting SWCNTs with specific chiral structures can mimic the required complex refractive index characteristics. Semiconducting SWCNTs exhibit strong optical absorption with steep cutoff arising from their quasi-one-dimensional excitons,[50–54] which are stable even at extremely high temperatures.[55–57] Thus, they can be used in optical devices operating at high temperatures. SWCNTs are rolled-up graphene sheets with different chiral structures (*i.e.*, chirality) depending on the roll-up vector described by a pair of chiral indices (*n, m*) (the chiral structure of SWCNTs is described in Supplementary Note 4 and Figure S5). Many properties of SWCNTs strongly depend on (*n, m*), particularly their optical resonance energy and optical gap.[58–63] As-synthesized SWCNTs contain many (*n, m*) species whose resonance absorption peaks overlap with each other; therefore, SWCNTs usually appear in black and are used as absorbing materials including solar absorber.[64–68] However, recent progress in structure separation techniques have enabled the fabrication of SWCNT membranes enriched with only one (*n, m*) species.[63,69–71] Figure 3b shows the complex refractive index spectra of (10,3) and (6,5) SWCNT membranes reproduced using the previously reported empirical formula.[63] These spectra reveal sharp exciton resonances in the near-infrared region, without any resonant absorption below the exciton resonance. The value of $\kappa$, except for the wavelength of the exciton resonance, is close to $\kappa \approx 2/\pi$, and $n$ increases with $\lambda$; these values are relatively close to the desired $n$ and $\kappa$, compared with those of other semiconductors. Unlike typical black mixtures, these membranes exhibit characteristic colors arising from their structure-dependent absorption (inset of Figure 3b). Because the exciton resonance wavelength is approximately proportional to the diameter of nanotubes, the optical properties of a SWCNT membrane can be tailored by appropriately selecting and combining specific (*n, m*) structures.

Figure 3b shows that membranes with only an (*n, m*) structure still fail to simultaneously achieve the desired $n$ and $\kappa$ across a wide spectral range. However, by mixing SWCNTs with multiple (*n,* $m$) structures, a material satisfying $n \approx \lambda/4d$ and $\kappa \approx 2/\pi$ in a wide spectral range can be obtained. Here, the complex refractive index spectrum for the in-plane response of a membrane with multiple SWCNT structures is considered as $\tilde{n} = \sqrt{1 + \tilde{\chi}_{mix}}$, where $\tilde{\chi}_{Mix}$ is the complex in-plane optical susceptibility of a mixed SWCNT membrane. $\tilde{\chi}_{mix}$ is predicted as the sum of the susceptibility of each ($n, m$) structure *via* mean field approximation:[72]

$$\tilde{\chi}_{mix} = \sum_i f_i \tilde{\chi}_i, \quad (1)$$

where $f_i$ and $\tilde{\chi}_i$ are the volume fraction and complex optical susceptibility of each ($n, m$) structures, respectively. For the in-plane response of SWCNT-mixture membranes, this approximation is justified because the depolarizing coefficient $L$ along axis of the nanotube can be approximated as $L = 0$ for the extremely high aspect ratio of SWCNTs. The out-of-plane response must be considered while dealing with incident light from various directions. In these membranes, SWCNTs are aligned within the membrane plane, and the in-plane response is primarily dominated by the excitations parallel to the nanotube axis.[73] Meanwhile, the out-of-plane response is mainly dominated by excitations perpendicular to the nanotube axis, making the SWCNT membranes birefringent. Because the variation in the complex refractive index of out-of-plane response $\tilde{n}_{out}$ is relatively small within the wavelength range of interest,[73] we consider it as a constant of 1.9 regardless of ($n, m$), regardless of the number of the mixed structures and their volume fractions for simplicity. This approximation well-reproduced the experimental results, which will be discussed in subsequent sections.

We first consider the five ($n, m$) structures ((10,3), (6,5), (9,4), (9,2), and (8,3)) whose complex refractive index spectra were reported in previous research[63] (Supplementary Note 5). Figure 3c presents a designed complex refractive index spectrum obtained by combining three typical ($n, m$) structures with an optimized mixture ratio which give rise to highest solar absorptance and absorber efficiency among the possible combinations of these five structures, as shown in Figure 3d. Compared with the single (*n, m*) cases (Figure 3b), the mixture more closely reproduced $\kappa \approx 2/\pi$ and $n \approx \lambda/4d$ over a broad wavelength range below the cutoff. The bottom panel of Figure 3c shows the calculated absorptance spectra of the optimized SWCNT membrane on a gold substrate. It is confirmed that an excellent spectral selectivity can be obtained *via* complex refractive index engineering by mixing appropriate (*n, m*) structures of semiconducting SWCNTs. Supplementary Note 5, and Figures S6 and S7 present the engineered complex refractive index and absorptance spectra obtained using two to five (*n, m*) structures, respectively. Within these five structures, there is no positive effect on solar absorptance and absorber efficiency when more than three structures are present. This indicates the importance of selecting the appropriate (*n, m*) structures. Using the reported empirical formula[63] and exciton energies of individual SWCNTs,[61] the structure selection can be extended to a wider range, enabling higher solar absorptance and absorber efficiency, and tailorable cutoff with respect to different operation temperature. Figure S8 presents an example of SWCNT membrane comprised five (*n, m*) structures including larger-diameter SWCNTs. The solar absorptance/infrared emittance ratio reaches 0.90/0.02 (Figure 3d), which is among the highest compared with planar dielectric–metal tandem structures reported in the literature.[29–38]

**Demonstration of carbon nanotube-based selective absorber.** To verify the concept discussed above, complex refractive index engineering was conducted using two representative SWCNT species, (10,3) and (6,5). These structures were used considering the availability of large amounts of starting chirality-sorted materials suitable for the experiment.[70] Figure 4a shows the main procedures of the sample fabrication. The SWCNT membranes were fabricated by the vacuum filtration of SWCNT dispersion with controlled components.[63,74] The thickness of the membranes was controlled based on the volume and concentration of SWCNT dispersion. Next, the fabricated membranes were transferred to sapphire substrate and gold-coated silicon substrate (hereafter, SWCNT-SSA) *via* a wet transfer process and subjected to 15-min vacuum annealing at 300 ºC.

Because the complex refractive index spectra of the membranes may slightly depend on the quality of the starting materials, dispersion, and fabrication, the complex refractive index spectra of (6,5)-enriched and (10,3)-enriched membranes used in this study were obtained (Figure S9), using a previously proposed method.[63] Subsequently, the optical spectra of the mixed SWCNT–gold tandem structure with varying thickness were simulated to numerically determine the combinations exhibiting the highest solar absorptance/infrared emittance ratio (Supplementary Note 5 and Figure S10). Notably, the highest ratio could be achieved when the chirality volume ratio (10,3):(6,5) was approximately 70%–80%:30%–20% and the thickness of the SWCNT membrane was approximately 100 nm. Figure 4b shows the complex refractive index spectra of the fabricated (10,3)–(6,5) mixed SWCNT membrane (Supplementary Note 5 and Figure S11) that agrees well with the predicted spectra. An additional peak is observed around 1.6 µm above the resonance wavelength of (10,3) SWCNTs, which is missing in the simulated spectrum shown in Figure 3. This peak is attributed to the response of SWCNT structures that were not completely removed during the separation of (10,3) SWCNTs. An increase in $\kappa$ at longer wavelength was deemed the Drude-like response of free carriers in residual metallic SWCNTs or large-diameter semiconducting SWCNTs unintentionally doped during separation. Nevertheless, the influence of finite $\kappa$ in the infrared region can be neglected for the proposed absorber design because the membrane is considerably thinner than the infrared wavelength and the response of the highly reflective substrate dominated the overall infrared response of the absorber (Supplementary Notes 2 and 5, and Figures S10 and S11).

Figure 4c shows the absorptance spectrum of SWCNT-SSA under normal incidence, reasonably agreeing with the simulated spectrum (dashed curve). SWCNT-SSA shows a high solar absorptance and low effective emittance of 0.84 and 0.03 (at 300 °C), respectively. The inset shows a typical optical image of SWCNT-SSA. Contrary to the reflective Au surface, the SWCNT-covered area appears black because of the high absorptance in the visible range. Figure 4d shows a map of the calculated power dissipation density (PDD) for normal incidence as a function of the wavelength and the position in the membrane, in which 0 nm is set at the interface of the SWCNT membrane and Au substrate. PDD is the energy dissipation per unit volume, defined as $P = \frac{1}{2}\varepsilon_0\varepsilon_2\omega|\boldsymbol{E}|^2$, where $\varepsilon_0$ is the permittivity of vacuum, $\varepsilon_2 = 2n\kappa$ is the imaginary part of the dielectric function, $\omega$ is the angular frequency, and $\boldsymbol{E}$ is the optical electric field. This indicates that power dissipation primarily occurs near the absorber surface, which is advantageous for localizing solar thermal energy in a limited space near the surface to generate high temperatures.

Figure 4e shows the absorptance spectra obtained at various incident angles using a laboratory-made setup and unpolarized light. The inset in Figure 4e shows the spectra simulated for each incident angles. The corresponding solar absorptance is plotted in Figure 4f as a function of the incident angle. The absorptance slightly decreases with increasing incident angle, and high sunlight absorptance maintained over various incident angles up to 60°. Furthermore, the SWCNT-SSA shows consistent cutoff wavelength across various incident angles, contrary to that observed in conventional interference-based absorbers.[46,75]

**Photothermal performance evaluation.** Figure 5a shows the experiment setup constructed for evaluating the SWCNT-SSA performance. Solar-thermal experiments were conducted in a vacuum chamber at room temperature (21 °C) under nonconcentrated light (1 SUN = 1 kW m$^{-2}$) emitted from a solar simulator. The left panel of Figure 5b shows an optical image of the SWCNT-SSA, wherein a K-type thermocouple was pasted on the SSA surface to measure and record time-dependent temperature curves under solar illumination. The right panel of Figure 5b presents an infrared image of SWCNT-SSA captured during the solar-thermal experiment, recorded using a thermal camera that can detect infrared radiation of 7–14 μm, indicating its low thermal radiation and emissivity. Figure 5c shows the temperature evolution of SWCNT-SSAs comprising mixed (10,3)–(6,5) SWCNTs, compared with that of SSAs with a similar structure but fabricated using only (10,3) or (6,5), and a quasi-blackbody absorber. From Figure 5c, the mixed SWCNT-SSA achieves the highest temperature of 170 °C under 1 SUN illumination, whereas that of the blackbody-like absorber reaches only 95 °C. The higher equilibrium temperature of the mixed SWCNT-SSA under the identical experimental conditions is ascribed to its considerably enhanced spectral selectivity. The equilibrium temperatures of (10,3) and (6,5) SSAs reach 161 °C and 132 °C, respectively. These results are reasonable because the (10,3) and (6,5) SSAs have solar absorptances/infrared emittances of 0.82/0.04 and 0.72/0.04, respectively, which are lower than that of the mixed membranes (0.84/0.03). These findings validate the proposed concept of tailoring the complex refractive index spectra by mixing multiple semiconducting SWCNT structures to realize a broadband SSA with a simple Dallenbach-type bilayer structure.

The as-fabricated SWCNT-SSA achieved high temperatures and efficiency. Although the (10,3)–(6,5) mixed SWCNT SSA could achieve a high solar absorptance/infrared emittance of 0.84/0.03, this value can be further optimized to 0.90/0.02 by adding other (*n, m*) structures, and improving the absorber spectrum. Because a small absorber was used (1 cm$^2$), the highest achievable temperature was limited by energy leakage to the measurement system. The theoretical upper limit of the blackbody-like absorber under 1 SUN illumination was 125 °C, derived based on radiative equilibrium with the environment (black dashed line in Figure 5c); however, the blackbody-like absorber achieved an equilibrium temperature of only 95 °C. The upper limit of the equilibrium temperature of the (10,3)–(6,5) mixed SWCNT-SSA under the radiative equilibrium was 480 °C in a similar calculation, considerably higher than the 170 °C. To reduce the system effect, the thermocouple was cut to reduce the conductive loss, leaving only the adhesive tape as the measurement point of the infrared camera (Supplementary Note 6 and Figures S12 and S13). Under 1 SUN illumination, the blackbody-like absorber achieved a higher equilibrium temperature of 100 °C, and the SWCNT-SSA achieved an even higher equilibrium temperature of 190 °C (Figure S14), indicating the potential of SWCNT-SSA to achieve higher equilibrium temperature. Figure 5d shows the equilibrium temperatures of the blackbody-like absorber and the (10,3)–(6,5) mixed SWCNT-SSA measured as a function of the solar power densities. Although the temperature of the blackbody-like absorber reached 150 °C under the 2 SUN illumination (2 kW m$^{-2}$) (increase of 50 °C compared with the 1 SUN illumination), the (10,3)–(6,5) mixed SWCNT-SSA reached 270 °C, showing an increase of 100 °C. The equilibrium temperature of the SWCNT-SSA under 1 SUN condition (190 °C) was considerably higher than that of the blackbody-like absorber under 2 SUN illumination, further confirming the potential of the mixed SWCNT-SSA.

# CONCLUSION

In this study, we developed a high-performance bilayer SSA based on the concept of Dallenbach-type absorber, using a SWCNT-based absorbing layer with a tailored complex refractive index spectrum. By mixing multiple types of semiconducting SWCNTs to approximate the theoretically required complex refractive index spectrum for to broadband solar selectivity, we demonstrated that high solar absorptance and low infrared emittance, with a low sensitivity to the angle of incidence, can be achieved. The proposed concept is different from the conventional method that simply combines a relatively thick absorptive material with a reflective substrate to extend the optical path length. To verify this, we experimentally engineered a complex refractive index spectrum using two representative SWCNT species, (10,3) and (6,5). The fabricated proof-of-concept SWCNT-SSA exhibited excellent spectral selectivity, with a solar absorptance of 0.84 and an infrared emittance of 0.03 (for 300 °C radiation); these values were maintained even at large incident angles. Spectral selectivity can be further improved to higher value of 0.90/0.02 if structure-sorted larger-diameter SWCNTs are available for the spectral engineering. Under non-concentrated and 2× concentrated sunlight, the SWCNT-SSA achieved equilibrium temperatures of approximately 190 °C and 270 °C, respectively, considerably surpassing those of a blackbody-like absorber (100 °C and 150 °C). Furthermore, our results clearly demonstrated the capability of designing dielectrics with desired optical properties achieved by mixing various structure-sorted SWCNTs, which can be employed for designing various types of SWCNT-based photonic devices beyond SSAs.

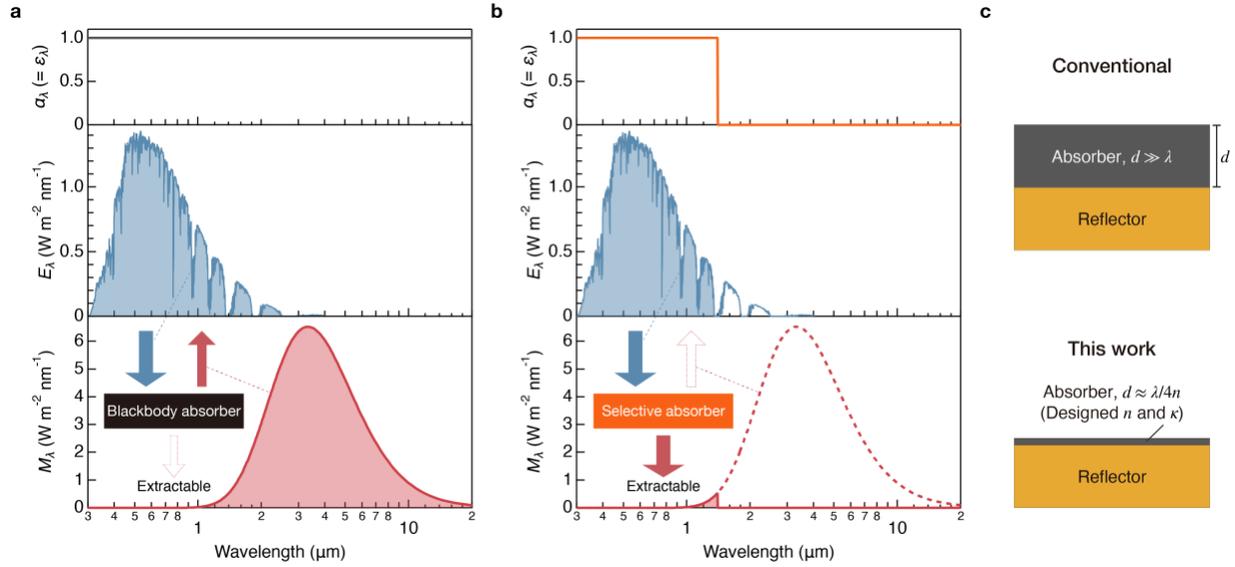

**Figure 1.** Fundamental design principle of the solar absorber. Solar-thermal conversion using (a) blackbody and (b) spectrally selective absorbers. Gray and orange curves in the top panels of (a) and (b) denote the absorptance $\alpha_\lambda$ (and emittance $\varepsilon_\lambda$) feature of the blackbody and spectrally selective absorber (SSA), respectively. The cutoff wavelength of SSA is set at 1.4 μm (Supplementary Note 1 and Figures S1 and S2). Blue curves in the middle panels denote the spectral irradiance $E_\lambda$ of AM 1.5 sunlight, and blue-filled regions represent the amount of sunlight absorbed. Red curves in the bottom panels represent the spectral radiant exitance $M_\lambda$ of a blackbody at 873 K, and red-filled regions represent thermal radiation. Arrows in the insets indicate sunlight input, thermal radiation leak, and extractable energy. (c) Conventional bilayer SSA and the proposed Dallenbach-type structure.

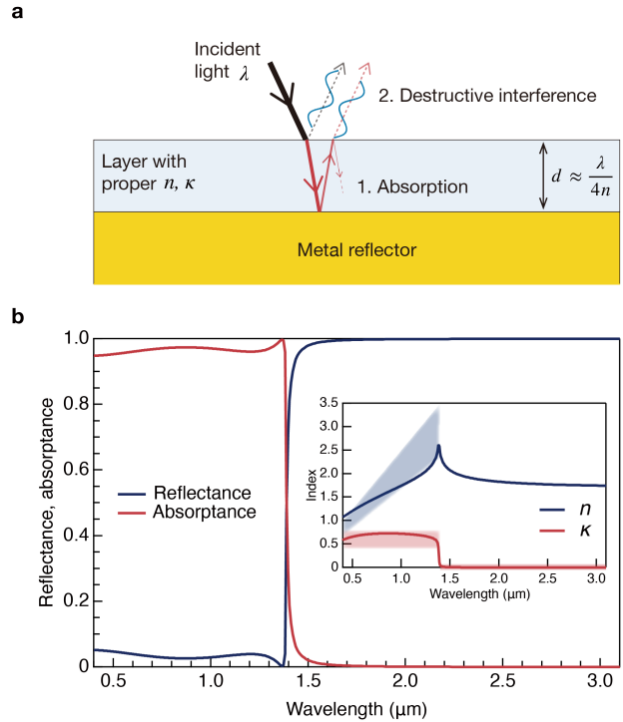

**Figure 2.** Design of an ideal dielectric for a bilayer SSA. (a) Principle for achieving full absorption at wavelength λ. For a given thickness *d*, an appropriate pair of refractive index *n* and extinction coefficient *κ* enable the combination of intrinsic absorption and destructive interference, resulting in a near-zero reflection. This can be extended to a wider wavelength range with appropriate curves of $n(\lambda)$ and $\kappa(\lambda)$. (b) Calculated spectra of the structure with hypothetical dielectric placed on a hypothetical lossless metallic surface. The thickness of the dielectric is 110 nm. Inset shows the artificial complex refractive index spectra of the hypothetical absorbing layer. Blue- and red-filled regions represent the appropriate value of *n* for layer thickness of 100–150 nm and $\kappa = 2/\pi$ with some tolerance.

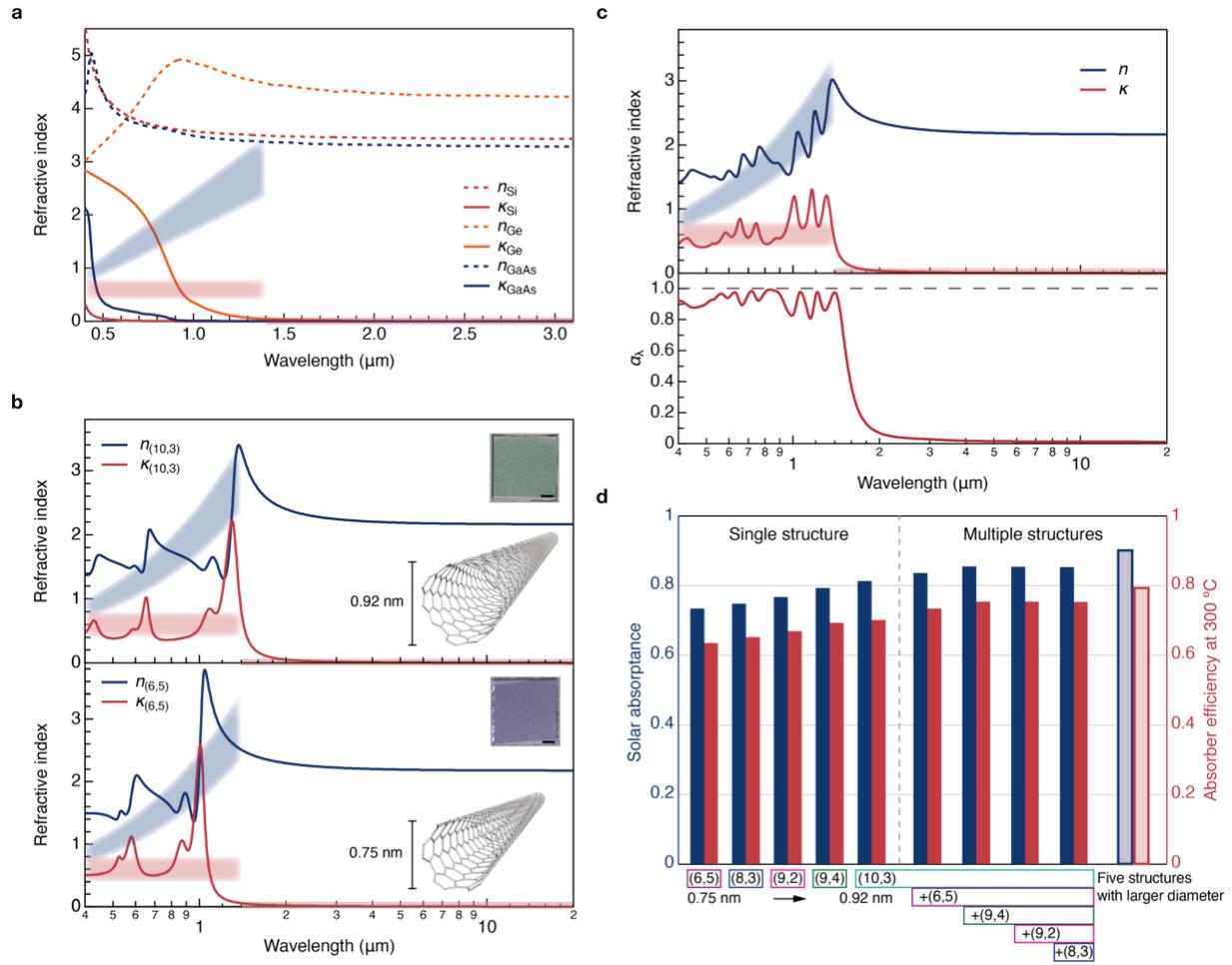

**Figure 3.** Complex refractive index in the target range. (a) Complex refractive index spectra of conventional semiconductors. Blue- and red-filled regions denote the desired value of $n$ for layer of thickness 100–150 nm and $\kappa = 2/\pi$ with some tolerance. (b) Complex refractive index spectrum of the membrane enriched with (upper) (10,3) SWCNTs and (bottom) (6,5) SWCNTs. Insets show the schematic of the (10,3) SWCNT, photograph of (10,3)-enriched membrane on sapphire, schematic of (6,5) SWCNT, and photograph of (6,5)-enriched membrane on sapphire. The scale bar is 1 mm. SWCNT schematics are drawn using VESTA 3.[76] (c) (Upper) complex refractive index spectra of SWCNT membrane comprising (10,3), (6,5), and (9,4), with the ratios of 50%, 20% and 30%, respectively. (Bottom) Calculated absorptance spectrum of the SWCNT–gold tandem structure. SWCNT membrane thickness is 105 nm. (d) Maximum of solar absorptance

$\alpha_{max}$ (blue bar charts) and absorber efficiency $\eta_{Max}$ at 300 °C (red bar charts) that can be reached when using the five reported chiral structures and their combinations from two to five structures. The two bar charts on the far right show the case using the other five chiral structures ((10,9), (11,4), (10,5), (9,2) and (7,5)). These structures have the diameters ranging from 0.82 nm to 1.29 nm.

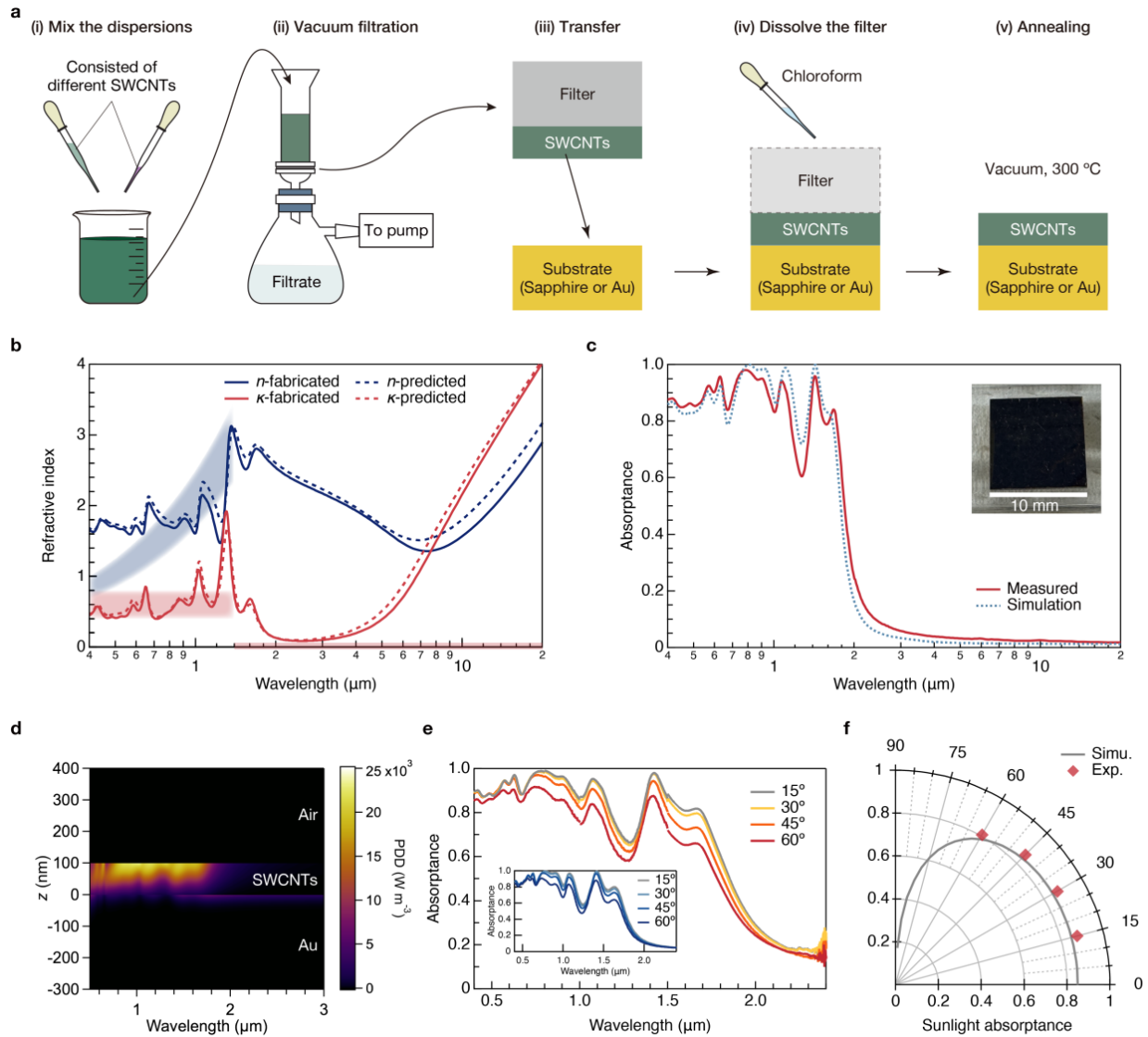

**Figure 4.** Demonstration of SWCNT-SSA. (a) Schematic of SWCNT membrane fabrication. (b) Determined complex refractive index spectra of the engineered SWCNT membrane compared with the one predicted from (10,3)-enriched and (6,5)-enriched membranes. The blue- and red-filled regions represent the desired value of *n* for layer thickness of 100–150 nm and $\kappa = 2/\pi$ with some tolerance. (c) Measured absorptance spectrum of fabricated SWCNT-SSA, compared with the simulation results. Inset shows a typical photograph of SWCNT membrane placed on the gold-coated substrate. (d) Calculated power dissipation density (PDD) within the SWCNT-SSA for normal incidence. (e) Absorptance spectra measured at various incident angles. Inset shows the simulated results. (f) Solar absorptance with respect to the angle of incidence, calculated from simulated and measured spectra.

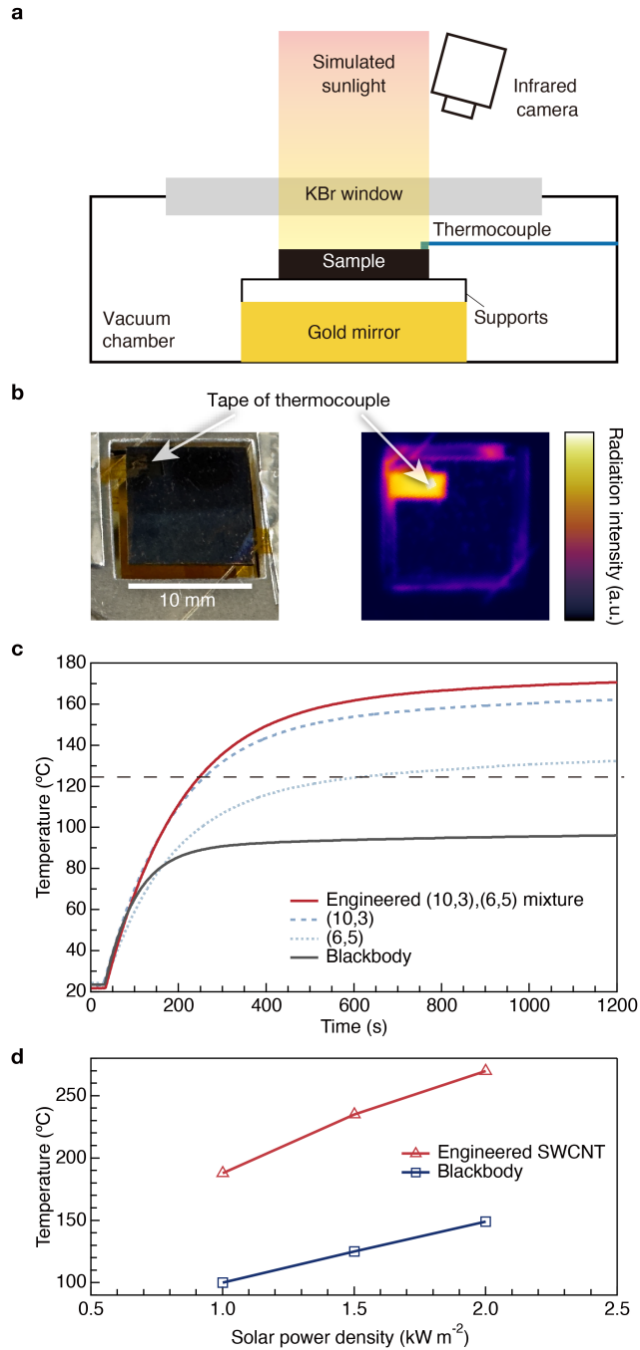

**Figure 5.** Solar-thermal performance of the designed SWCNT-SSA. (a) Schematic of the experiment setup to measure the temperature response under sunlight illumination. (b) (left) Photograph of the designed SWCNT-SSA and (right) infrared image during experiment. At high temperatures, SWCNT-SSA shows weak infrared radiation. Bright region is the thermal radiation

from the thermocouple tape. (c) Temperature response under nonconcentrated sunlight illumination. Black-dashed line shows the theoretical upper limit of the blackbody under nonconcentrated sunlight illumination. (d) Equilibrium temperature of the absorbers under different solar intensities (1, 1.5, 2 kW m$^{-2}$), with further suppressed systematic effect.

# METHODS

**Simulation and optical design.** A multiple reflection model based on Fresnel's equations was used for spectrum simulations.[77] At normal incidence, there is no difference between *s*-polarized light and *p*-polarized light. The total reflectance is expressed as:

$$R = \left| \frac{\tilde{r}_{AC} + \tilde{r}_{CS} \exp(-i2\tilde{\beta})}{1 + \tilde{r}_{AC}\tilde{r}_{CS} \exp(-i2\tilde{\beta})} \right|^2, \quad (2)$$

where $\tilde{r}_{AC}$ and $\tilde{r}_{CS}$ denote the amplitude reflection coefficients at the air–SWCNT and SWCNT–substrate interfaces, respectively. $\tilde{\beta}$ is the absorption-induced phase variation during propagation in the SWCNT layer, expressed as follows:

$$\tilde{r}_{AC} = \frac{1 - \tilde{n}_{in}}{1 + \tilde{n}_{in}}, \quad (3)$$

$$\tilde{r}_{CS} = \frac{\tilde{n}_{in} - \tilde{n}_{sub}}{\tilde{n}_{in} + \tilde{n}_{sub}}, \quad (4)$$

$$\tilde{\beta} = \frac{2\pi \tilde{n}_{in} d}{\lambda}, \quad (5)$$

where $\tilde{n}_{in}$ is the in-plane complex refractive index of the SWCNT membrane, $\lambda$ is the wavelength, $d$ is the thickness of the SWCNT membrane, and $\tilde{n}_{Sub}$ is the complex refractive index of the substrate. Because the substrate is a substantially thick metal that prevents transmission, the absorptance of the structure can be expressed as $A = 1 - R$.

It should be noted that SWCNT membranes are birefringent, with the out-of-plane complex refractive index $\tilde{n}_{out}$ differing from the in-plane one $\tilde{n}_{in}$.[73] Consequently, the expression for reflectance becomes more complicated while considering light incident from various angles:

$$R^s(\tilde{n}_{in}, \tilde{n}_{sub}, \lambda, \theta, d) = \left| \frac{r^s_{AC}(\tilde{n}_{in}, \theta) + r^s_{CS}(\tilde{n}_{in}, \tilde{n}_{sub}, \theta) \exp(-i2\beta^s(\tilde{n}_{in}, \lambda, \theta, d))}{1 + r^s_{AC}(\tilde{n}_{in}, \theta) r^s_{CS}(\tilde{n}_{in}, \tilde{n}_{sub}, \theta) \exp(-i2\beta^s(\tilde{n}_{in}, \lambda, \theta, d))} \right|^2, \quad (6)$$

and

$$R^p(\tilde{n}_{\text{in}}, \tilde{n}_{\text{out}}, \tilde{n}_{\text{sub}}, \lambda, \theta, d) = \left| \frac{r_{\text{AC}}^p(\tilde{n}_{\text{in}}, \tilde{n}_{\text{out}}, \theta) + r_{\text{CS}}^p(\tilde{n}_{\text{in}}, \tilde{n}_{\text{out}}, \tilde{n}_{\text{sub}}, \theta) \exp(-i2\beta^p(\tilde{n}_{\text{in}}, \tilde{n}_{\text{out}}, \lambda, \theta, d))}{1 + r_{\text{AC}}^p(\tilde{n}_{\text{in}}, \tilde{n}_{\text{out}}, \theta) r_{\text{CS}}^p(\tilde{n}_{\text{in}}, \tilde{n}_{\text{out}}, \tilde{n}_{\text{Sub}}, \theta) \exp(-i2\beta^p(\tilde{n}_{\text{in}}, \tilde{n}_{\text{out}}, \lambda, \theta, d))} \right|^2, \quad (7)$$

for *s*- and *p*-polarized light, respectively (Supplementary Note 7). The reflectance of unpolarized light is obtained by averaging of the responses for *s*- and *p*-polarized lights.

**Sample fabrication.** Single-chirality SWCNTs were separated from the starting materials, CoMoCAT (SG65, Sigma-Aldrich) and HiPCo (NanoIntegris) nanotubes, *via* gel column chromatography.[70] The separated SWCNTs were dispersed in an aqueous surfactant solution containing sodium cholate (SC), sodium dodecyl sulfate (SDS), and sodium deoxycholate (DOC). The surfactant concentrations were 0.5 % SC, 1.0 % SDS, and 0.155% DOC for (10,3) dispersions and 0.5 %SC, 0.5% SDS, and 0.028% DOC for (6,5) dispersions. The concentration and amount of SWCNTs in the dispersion were evaluated from their $S_{11}$ resonance peaks (Figure S15).[78] Engineered SWCNT dispersion was prepared by mixing the single-chirality SWCNT dispersions in the designed ratio.

The SWCNT membranes were fabricated *via* the vacuum filtration method,[63,74] and their thickness was controlled to be approximately 50 nm by adjusting the concentration and volume of SWCNT dispersions. First, the SWCNT dispersion was preliminarily filtered using a hydrophobic polytetrafluoroethylene syringe filters (with pore size of 5 μm). Then, the dispersion was diluted to reduce the surfactant concentration below their critical micelle concentration and poured into a vacuum filtration system. During the filtration, the solvent passed through the filter while suspended SWCNTs were captured by a 100-nm-pore polycarbonate (PC) membrane filter (MERCK, VCTP02500), forming a network of SWCNTs. The pressure applied to the system was monitored by a pressure gauge, and the average filtration speed was approximately 0.5 mL min$^{-1}$. The remnant surfactants on the fabricated membrane were removed by washing with 80 °C hot water and air-drying for 30–60 mins under a pressure of approximately 5 kPa. The filtrate is collected and filtered again to retain as many SWCNT as possible on the filter paper.

To fabricate the SWCNT-SSA, approximately 100-nm thick gold layer was coated on a 1-cm$^2$ silicon substrate *via* electron-beam evaporation. The SWCNT membrane on the PC filter was cut to the same size and placed on the substrate with the filter side facing up. Next, chloroform was dropped on the surface to dissolve the PC filter and release the SWCNT membrane. After the spontaneous evaporation of chloroform, the SWCNT membrane adhered to the substrate, followed by washing with fresh chloroform, ethanol, and acetone. Then, the sample was annealed under vacuum ($10^{-7}$–$10^{-6}$ kPa) at 300 °C for 15 min. The SWCNT-SSA with a thickness of approximately 100 nm was fabricated by stacking two SWCNT membranes and repeating the aforementioned procedures twice. A blackbody-like absorber was prepared by spraying the blackbody coating (Japansensor Corporation JSC-3) on the same silicon substrate for comparison. This absorber was annealed in open air at 200 °C for 20 minutes.

**Optical measurement.** The optical spectra at normal incidence were measured using an ultraviolet–visible–near infrared (UV–Vis–NIR) spectrometer (V-770, JASCO) and a Fourier transform infrared (FTIR) spectrometer (FT/IR-6600, JASCO). The UV–Vis–NIR and FTIR spectrometers covered wavelengths of 0.2–2.5 μm (0.496–6.20 eV) and 0.7–20 μm (0.06–1.77 eV), respectively. Unpolarized light sources were used for measurements, and the incident angles were 5° and 10° for reflectance measurements in the UV–Vis–NIR and FTIR measurements, respectively.

The reflectance ($R$) spectra at various incident angles were measured using a laboratory-made optical setup. In the incident part, broadband light from a halogen lamp passing through an optical fiber was collimated by a parabolic mirror, and the beam diameter was sized using an iris diaphragm. The detection part comprised a long-pass filter, an iris diaphragm, and a focusing lens. The long-pass filter was chosen based on the measurement range to eliminate the effect of second-order diffraction. The reflected light was focused by the lens onto the optical fiber connected to the detector. Several detectors (Ocean Optics USB2000+, 0.4–1.1 μm; Ocean Photonics NIRQuest+, 0.9–1.7 μm; ARCoptix FTNIR-L1-025-2TE, 0.9–2.5 μm) were combined for a broadband range. The spectra were obtained by changing the position of the incident and detection parts using automatic rotating stages. The absorptance ($A$) spectrum was calculated as $A = 1 - R$.

**Calculation of power dissipation density.** An open-source simulation tool based on rigorous coupled-wave analysis method from GitHub was used to calculate the PDD.[79–82] The complex refractive index spectrum of gold was modeled using the Lorentz–Drude model.[83]

**Solar-thermal measurement.** The solar-thermal performance was evaluated using a designed chamber equipped with a KBr window and a solar simulator (Asahi Spectra HAL-320W). The KBr window was selected owing to its high and uniform transmittance over the relevant wavelength of 0.3–20 μm, allowing the sunlight and thermal radiation to pass without changing the spectral shapes. The solar flux was calibrated to 1 kW m$^{-2}$ (1 SUN) using the attached standard silicon solar cells and thermopiles (Thorlabs PM100D Power meter Console, S425C Power sensor head). A tape-attached K-type thermocouple (RKC INSTRUMENT INC. ST-51S) was pasted on the front side of the sample for temperature measurement and connected to a data recorder (RKC INSTRUMENT INC. VGR-B100). The temperature was also measured by an infrared camera (ARGO Corporation Xi400) using the thermal radiation from the thermocouple tape. The recording rate was set as 1 Hz for the data recorder and infrared camera. To approximately evaluate the solar-thermal performance, the chamber was evacuated to $10^{-5}$–$10^{-4}$ kPa using a turbo pump during the experiment to reduce the thermal convection. The sample was suspended by two quartz fibers (50-μm diameter) to reduce contact area and the thermal conduction to the experiment setup. A gold mirror was placed on the backside of the absorber to reflect the thermal radiation back. The experiment time was 30 min for each sample; the light source was blocked for the first 30 s; then, the shutter was opened, and the sample was illuminated for approximately 20 min. After 20 min, the shutter was closed again.

ASSOCIATED CONTENT

**Supporting Information.** Supporting Information Available: Efficiency-related calculation; deviation of the ideal value of $\kappa$; spectrum simulation of hypothetical structures; definition of chiral indices ($n, m$) of SWCNTs; procedures of complex refractive index determination and absorber design; calculation of the upper limit of the equilibrium temperature; multiple reflection model for angle-dependent optical simulation; absorption spectra of single-chirality SWCNT dispersion.

AUTHOR INFORMATION

**Corresponding Author**

Yuhei Miyauchi – Institute of Advanced Energy, Kyoto University, Uji, Kyoto 611-0011, Japan; Department of Mechanical Engineering, The University of Tokyo, 7-3-1 Hongo, Bunkyo-ku, Tokyo 113-8656, Japan; https://orcid.org/0000-0002-0945-0265; Email: ymiyauchi@g.ecc.u-tokyo.ac.jp

**Author Contributions**

Y.M. conceived the concept and directed the project. T.T. and H.K. provided the raw dispersions. H.W. prepared the samples. H.W. and T.N. arranged the optical setup. T.N provided the framework of simulation program and H.W. conducted the programming. H.W. designed the solar-thermal setup. H.W conducted all the measurements. H.W., T.N. and Y.M. considered the data processing. M.H. performed the calculation of power dissipation density. H.W. prepared the first draft. All the authors contributed to writing of the manuscript.

**Notes**

The authors declare no competing financial interest.

ACKNOWLEDGMENT

This work was supported by JSPS KAKENHI Grant Numbers JP22K18287(Y.M.), JP24H00044(Y.M.), JP19K15384(T.N.), JP21K14486(T.N.), JP23H01791(T.N.), JP23KJ1353(H.W.), JP20H05668(H.K.), JP24K01282(T.T.), and JST CREST Grant Number JPMJCR185(Y.M.), and JST FOREST Grant Number JPMJFR222N(T.N.). We thank Atsushi Sakurai for discussion on the calculation procedure of PDD.

# Supplementary Information

**Supplementary Note 1:**

**Relation between the operating temperature and cutoff wavelength**

The dashed curves in Figure S1 show the blackbody radiation at various temperatures calculated based on Planck's law. As the operating temperature increases, the distribution of blackbody radiation shifts to shorter wavelengths. Therefore, a cutoff wavelength at shorter wavelengths is desired for suppressing thermal radiation at higher operating temperature.

**Overall conversion efficiency of solar-thermal conversion**

The overall conversion efficiency can be expressed as follows:[23]

$$\eta_{\text{total}} = \eta_{\text{Abs}} \eta_{\text{Carnot}}, \tag{S1}$$

where $\eta_{\text{Abs}}$ is the conversion efficiency of the absorber and $\eta_{\text{Carnot}}$ is the Carnot efficiency, which can be expanded as follows:

$$\eta_{\text{Abs}} = \frac{C \int d\lambda\, \alpha(\lambda) E_{\text{Solar}}(\lambda) - \int d\lambda\, \varepsilon(\lambda) M_{\text{BB}}(\lambda, T_{\text{Abs}})}{C \int d\lambda\, E_{\text{Solar}}(\lambda)}, \tag{S2}$$

$$\eta_{\text{Carnot}} = 1 - T_{\text{Amb}}/T_{\text{Abs}}, \tag{S3}$$

where $C$ is the solar concentration factor, $E_{\text{Solar}}(\lambda)$ is the solar irradiance (AM 1.5 standard spectrum[84] or blackbody radiation at 5800 K) at wavelength $\lambda$ and $M_{\text{BB}}(\lambda, T_{\text{Abs}})$ is the spectral radiant exitance of a blackbody at the operating temperature of the absorber $T_{\text{Abs}}$, calculated using Planck's law, and $T_{\text{Amb}}$ is the ambient temperature (298 K). $\alpha(\lambda)$ and $\varepsilon(\lambda)$ are the spectral absorptance and emittance of the absorber, respectively. Based on Kirchhoff's law of thermal radiation, we have $\varepsilon(\lambda) = \alpha(\lambda)$ under thermal equilibrium.

For the nonconcentrated case ($C = 1$), we calculated the maximum of $\eta_{total}$ as a function of cutoff wavelength $\lambda_{cut}$, assuming $\alpha(\lambda) = 1$ for $\lambda < \lambda_{cut}$ and $\alpha(\lambda) = 0$ for $\lambda > \lambda_{cut}$. As shown in Figure S2a, the maximum of $\eta_{total}$ under nonconcentrated light occurs at $\lambda_{cut} \approx 1.4$ μm. Figure S2b–d show the relation between operating temperature and each efficiency for absorbers with different cutoff wavelengths. As the operating temperature increases, $\eta_{Abs}$ decreases because of the enhanced and blue-shifted thermal radiation, whereas $\eta_{Carnot}$ increases with the operating temperature. Therefore, as the temperature increases, their product $\eta_{total}$ gradually increases and reaches a maximum value, after which it decreases to 0. For $\lambda_{cut} \approx 1.4$ μm and nonconcentrated sunlight, the maximum of total efficiency occurs at approximately 873 K. Furthermore, $\lambda_{cut}$ at longer wavelength corresponds to a lower operating temperature.

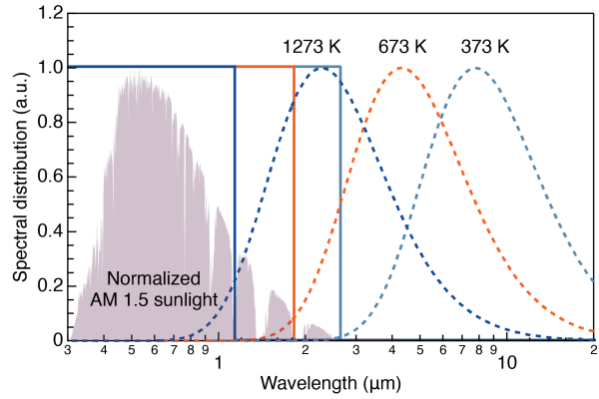

**Figure S1.** Spectral distribution of AM 1.5 sunlight (shadowed region) and blackbody radiation at varying temperatures (dashed curves). The solid curves represent the appropriate cutoff for each operating temperature.

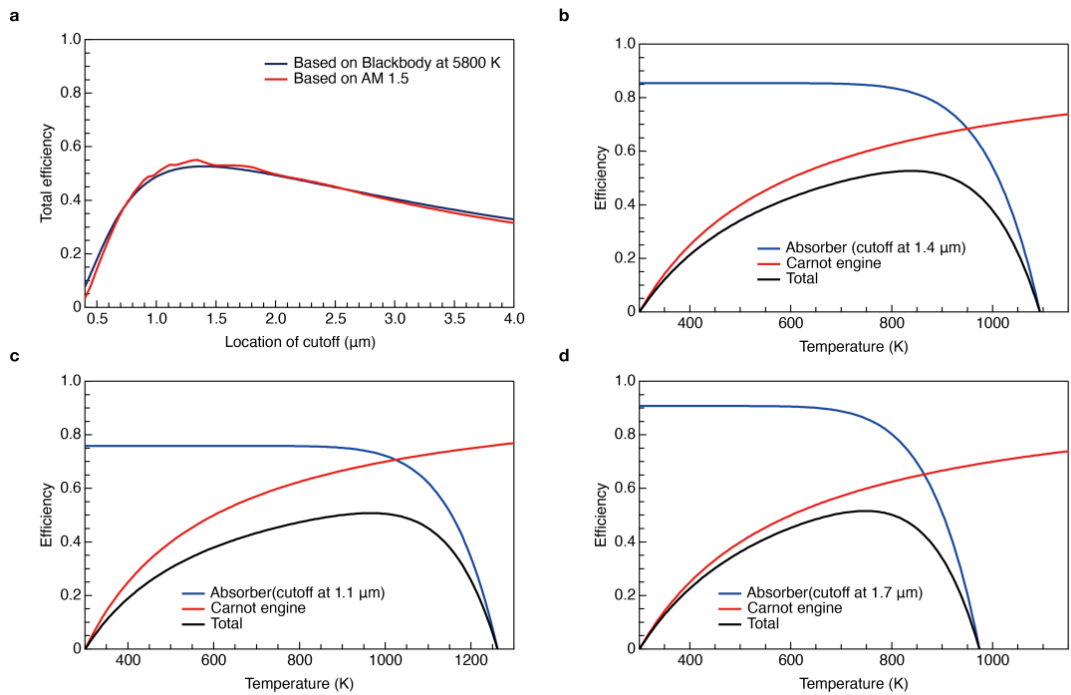

**Figure S2.** (a) Maximum of overall efficiency regarding the location of the cutoff wavelength. (b)–(d) Relation between operating temperature and efficiency for the spectral selective absorber (SSA) with a cutoff wavelength at (b) 1.4, (c) 1.1, and (d) 1.7 μm.

**Supplementary Note 2**

**Deviation of the ideal value of $\kappa$**

As mentioned in the main text, if a semiconductor with appropriate $n$ and $\kappa$ that fulfill $|r_1| \approx \exp(-\pi\kappa/n)$ at $\lambda$ is available, near-perfect absorption at this specific wavelength can be achieved. To extend it to a broader wavelength range, we consider a function $f[n(\lambda), \kappa(\lambda)] \equiv |r_1| - \exp(-\pi\kappa(\lambda)/n(\lambda))$ expresses the difference of the amplitude of the reflected field by the first and second surfaces. Herein, $f = 0$ indicates the optimal condition for reflection attenuation to maximize absorptance, and $\tilde{n}(\lambda) = n(\lambda) + \kappa(\lambda)$ is the complex refractive index spectrum. Figure S3 shows the absolute value of $|f|$ as a function of $n(\lambda)$ and $\kappa(\lambda)$. For $n(\lambda) > 1$, the value of $\kappa(\lambda)$ yielding near-zero $f$ is almost constant of $\approx 0.6$, regardless of $n(\lambda)$. This result can be analytically obtained. For a large $n(\lambda)$, $|r_1| \approx (n(\lambda) - 1)/(n(\lambda) + 1)$ and $\exp(-\pi\kappa(\lambda)/n(\lambda)) \approx 1 + (-\pi\kappa(\lambda)/n(\lambda))$; the approximate solution of $f = 0$ is therefore $\kappa(\lambda) \approx (2/\pi)(n(\lambda)/(n(\lambda) - 1))$. In the limit of $n(\lambda) \gg 1$, $\kappa(\lambda)$ approaches to constant $\kappa \approx 2/\pi \approx 0.64$, as is consistent with the result demonstrated in Figure S3. These are the major requirements for the imaginary part $\kappa(\lambda)$ in the solar spectral range from the ultraviolet to near-infrared range.

In the near-to-mid infrared region, $\kappa(\lambda) = 0$ is also required to suppress the absorption and avoid unwanted thermal radiation along with sharp transition from $\kappa(\lambda) \approx 0.6$ to 0 at the cutoff wavelength. Finally, with regard to the far infrared response where $\lambda \gg d$, there is almost no phase variation during the propagation in the membrane and phase variation-induced interference. Then, the reflectance can be expressed as follows:

$$R = \left|\frac{\tilde{r}_{AC} + \tilde{r}_{CS}}{1 + \tilde{r}_{AC}\tilde{r}_{CS}}\right|^2. \tag{S4}$$

For the metallic substrate, we have $|\tilde{n}_{\text{Sub}}| \gg |\tilde{n}_{\text{in}}|$. Therefore, $\tilde{r}_{\text{CS}} \approx -1$ and $R \approx |(\tilde{r}_{\text{AC}} - 1)/(1 - \tilde{r}_{\text{AC}})|^2 \approx 1$, indicating that reflection is dominated by the response of the metal substrate, and there is no strong requirement for κ.

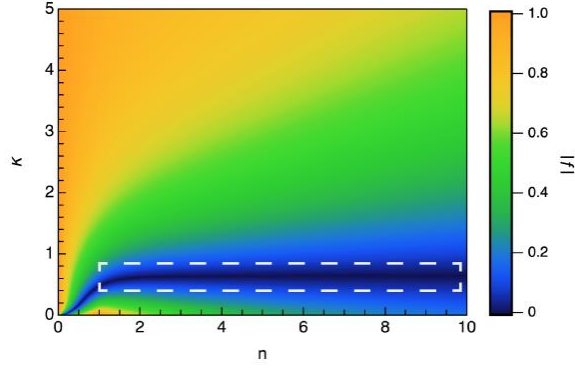

**Figure S3.** Absolute value of $|f|$ as a function of $n(\lambda)$ and $\kappa(\lambda)$. White dashed rectangle denotes the region with the minimum value.

**Supplementary Note 3**

**Spectrum simulation of hypothetical structures**

We calculate spectra of the structure in which a hypothetical absorbing layer is placed on a hypothetical lossless metallic surface.

As mentioned in the main text, the ideal absorber layer has the complex refractive index spectra that $\kappa(\lambda) \approx 0.6$ for $\lambda \leq 1.4$ μm and $\kappa(\lambda) = 0$ for $\lambda > 1.4$ μm, in addition to $n(\lambda) = \lambda/4d$ for a given thickness. Figure S4a shows the cases for thickness $d = 100$, 120 and 150 nm. Figure S4b shows the absorptance spectra calculated from these values. However, it should be noted that these hypothetical $n(\lambda)$ and $\kappa(\lambda)$ are not valid in physics because they are not dependent with each other. $n(\lambda)$ and $\kappa(\lambda)$ of a real material should satisfy Kramers-Kronig relation owing to causality. Herein, we constructed the complex refractive index spectrum satisfying the Kramers-Kronig relation by considering the summation of the responses of multiple Lorentz oscillators as follows:

$$\tilde{n}_{\text{Abs}} = \sqrt{1 + \sum_{n}^{N} A_n \exp(B_n)\, \tilde{\chi}_L^n(\omega)}, \qquad (S5)$$

where $A_n$ is a linear factor, $B_n$ is an exponential factor and $\tilde{\chi}_L^n$ is the $n$-th Lorentz oscillator. These are described as $A_n = \omega_L^n/\omega_{\text{cutoff}}$, $B_n = -3\omega_L^n/\omega_{\text{cutoff}}$, and $\tilde{\chi}_L^n(\omega) = f_L^n[(\omega_L^{n\,2} - \omega^2) - i\omega\gamma_L^n]^{-1}$, respectively. $\omega$ is the optical frequency; $\omega_{\text{cutoff}}$ is the cutoff frequency; and $f_L^n$, $\omega_L^n$, and $\gamma_L^n$ are the strength, resonant frequency, and damping term of the $n$-th Lorentz oscillator, respectively. Herein, $\hbar\omega_{\text{cutoff}}$, $f_L^n$, $\hbar\omega_L^n$ and $\hbar\gamma_L^n$ are set as 0.9 eV ($\approx$1380 nm), 0.01 g$^{-1}$ cm$^3$ eV$^2$, $(0.9 + 0.005n)$ eV, and 0.005 eV, respectively. $\hbar$ is the reduced Planck constant.

The lossless metallic surface was modeled using Drude model considering a perfector conductor case, where $\tilde{n}_{\text{PEC}} = \sqrt{1 - \omega_p^2/\omega^2}$. $\omega_p$ is the plasma frequency, and $\hbar\omega_p$ is set as 10 eV.

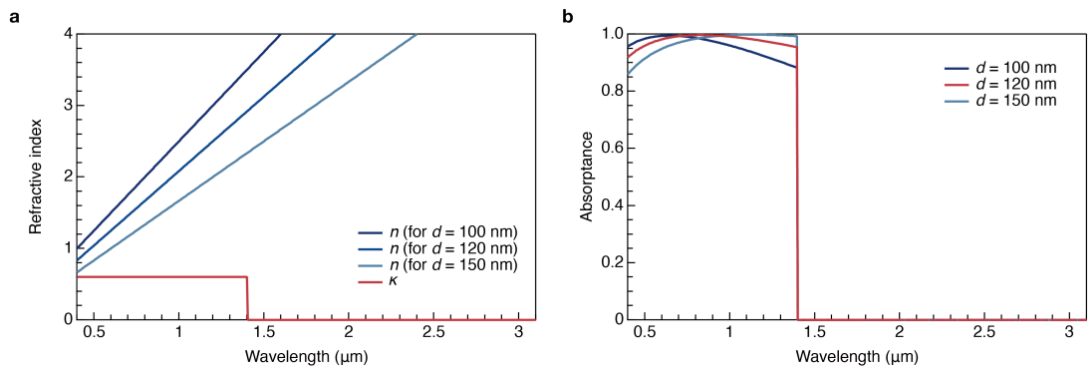

**Figure S4.** (a) Real part $n$ and imaginary part $\kappa$ of the hypothetical absorbing layer with varying thicknesses. (b) Absorptance spectra of the artificial structure simulated with the thicknesses of 100, 120 and 150 nm using the corresponded $n$.

**Supplementary Note 4**

**Definition of chiral indices (*n*, *m*) of SWCNTs**

Figure S5 shows a schematic to explain the definition of chiral indices. An SWCNT can be considered a rolled-up graphene sheet, in which $C_h$ is a vector denotes the two connected atoms. The chiral indices (*n*, *m*) denote the direction of $C_h$ concerning the graphene lattice basis vectors $a_1$ and $a_2$.

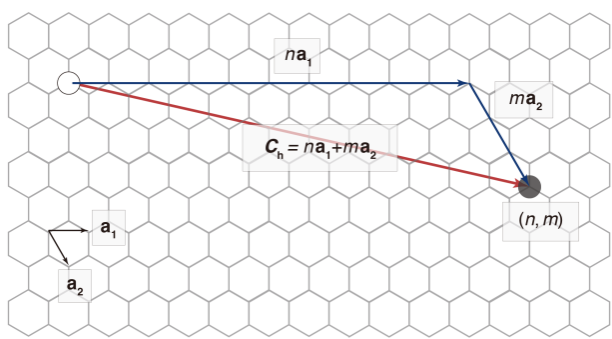

**Figure S5.** Typical schematic of chiral indices definition.

**Supplementary Note 5**

**Complex refractive index determination and absorber design**

The complex refractive index and optical susceptibility spectra of single-chirality SWCNT membranes were modeled using the previously reported empirical formula.[63,73] Then, we modeled the complex refractive index spectra of (10,3)–(6,5) SWCNT membranes, with the ratio ranging from 90%:10% to 10%:90%. Their real part and imaginary parts are shown in Figure S6a and S6b, respectively. For each mixture ratio, the optical spectra of the SWCNT–gold tandem structure with varying thickness were simulated. For each thickness, the sunlight absorptance ($\alpha^t$), effective emittance ($\varepsilon^t$) and absorber efficiency $\eta_{Abs}$ at 300 °C were calculated based on Eqs. (S6), (S7), and (S2), respectively.

$$\alpha^t = \frac{\int_{0.4\ \mu m}^{4\ \mu m} d\lambda \alpha(\lambda) E_{Solar}(\lambda)}{\int_{0.4\ \mu m}^{4\ \mu m} d\lambda E_{Solar}(\lambda)}, \quad (S6)$$

and

$$\varepsilon^t = \frac{\int_{0.4\ \mu m}^{20\ \mu m} d\lambda \varepsilon(\lambda) M_{BB}(\lambda, T)}{\int_{0.4\ \mu m}^{20\ \mu m} d\lambda M_{BB}(\lambda, T)}, \quad (S7)$$

Figure S6c shows the maximum of $\eta_{Abs}$ can be obtained with respect to the ratio of (10,3) in the mixture membrane. The highest value was obtained at the ratio of (10,3):(6,5)= 70%–80%:30%–20%. Figure S6d shows $\eta_{Abs}$ regarding the membrane thickness of the (10,3):(6,5)=70%:30% membrane. The highest value was achieved around $d = 110$ nm, consistent with the proposed concept.

Next, we model the complex refractive index spectra of the SWCNT membrane consisted of more (*n, m*) structures. In addition to (10,3) and (6,5), three additional chiral structures (9,4), (9,2) and (8,3) were selected using the parameters reported.[63] Among these five structures, the maximum of binary-mixed SWCNT membrane is achieved with the combination of (10,3)–(6,5); and the maximum of ternary- and quaternary-mixed SWCNT membranes is achieved with the combinations of (10,3)–(6,5)–(9,4) and (10,3)–(6,5)–(9,4)–(9,2), respectively. The real part and imaginary parts of their complex refractive index spectra are shown in Figure S7a and S7b. All the ratio combinations were calculated with a step length of 10% to determine the one with the highest $\eta_{Abs}$ at 300 °C. The complex refractive index spectra of the champion ratio for ternary-, quaternary-, and pentanary-mixed SWCNT membranes, and the simulated absorptance spectra are shown in Figure S7c–h. Within these five structures, there is no positive effect on solar absorptance and absorber efficiency when more than three structures are present. Notably, a higher sunlight absorptance and absorber efficiency can be obtained by adjusting the operating temperature and cutoff wavelength, selecting other (n, m) structures, and adding additional (n, m) structures. Here, we present an example using larger-diameter SWCNTs than the above five structures. Figure S8a and S8b present the complex refractive index spectra of (10,9), (11,4), (10,5), (9,2) and (7,5) SWCNTs, calculated using reported empirical formula[63] and exciton energies.[61] Figure S8c presents the complex refractive index spectrum of SWCNT membrane comprised of these five structures and Figure S8d presents the absorber efficiency at 300 °C of SWCNT–gold tandem structure with respect to the membrane thickness. The highest value was achieved around $d =$ 112 nm, almost the same value. The inset shows the absorptance spectrum of SWCNT–gold tandem with the membrane thickness of 112 nm, exhibiting an excellent spectral selectivity. The solar absorptance/infrared emittance ratio are calculated as 0.90/0.02.

To verify the proposed concept, we demonstrated the complex refractive index engineering using two representative nanotube species, (10,3) and (6,5). Because the complex refractive index spectrum may slightly vary by case (owing to the conditions of the starting materials, dispersion, and fabrication), we first determined it for the practical (6,5)-enriched and (10,3)-enriched membranes using a previously reported procedure.[63] In brief, we fabricated SWCNT membranes and transferred them onto a sapphire substrate and a stainless-steel washer (freestanding membrane), respectively. The complex refractive index spectra were modeled by superposing various oscillators representing the optical responses of SWCNTs. Using the modeled complex refractive index spectra, the reflectance and transmittance spectra were simulated and compared with the experimentally measured spectra. Finally, the complex refractive index spectrum of SWCNT membrane was determined as the one successfully reproduced the measured spectra. The optical spectra and determined complex refractive index spectra of (10,3) and (6,5) are shown in Figure S9 a-c and S9 e-g, respectively. Figure S9d and S9h show the absorptance spectra simulated using the determined complex refractive index spectra that considered and neglected the Drude response. The effect of Drude response was negligible when the thin membrane was placed on a metallic substrate; thus, the Drude response could be neglected during the simulation. Using the determined complex refractive index spectra of (10,3) and (6,5), the complex refractive index spectra of (10,3)–(6,5) SWCNT membranes were modeled using the same method, with the ratio ranging from 90%:10% to 10%:90%. Their real and imaginary parts are shown in Figure S10a and S10b. The optical spectra of the SWCNT–gold tandem structure with varying thickness were simulated for each mixture ratio. For each thickness, the absorber efficiency $\eta_{\text{Abs}}$, Carnot efficiency $\eta_{\text{Carnot}}$, and overall conversion efficiency $\eta_{\text{total}}$ at various temperature were simulated based on Eqs. (S1)-(S3). The maximum of $\eta_{\text{total}}$ that can be achieved is shown in Figure S10c with respect to the ratio of (10,3) in the membrane. The highest value was obtained at a ratio of (10,3):(6,5)=80%:20%, which is slightly different from the above simulation. Figure S10d shows $\eta_{\text{Total}}$ regarding the membrane thickness of the 80%:20% membrane. The highest value was obtained at approximately $d = 100$ nm. Based on the simulation result, the ratio of (10,3):(6,5)=80%:20% was chosen for SSA fabrication. The optical spectra of the prepared mixture membrane were also measured, and the complex refractive index spectrum were determined for comparison (Figure S11).

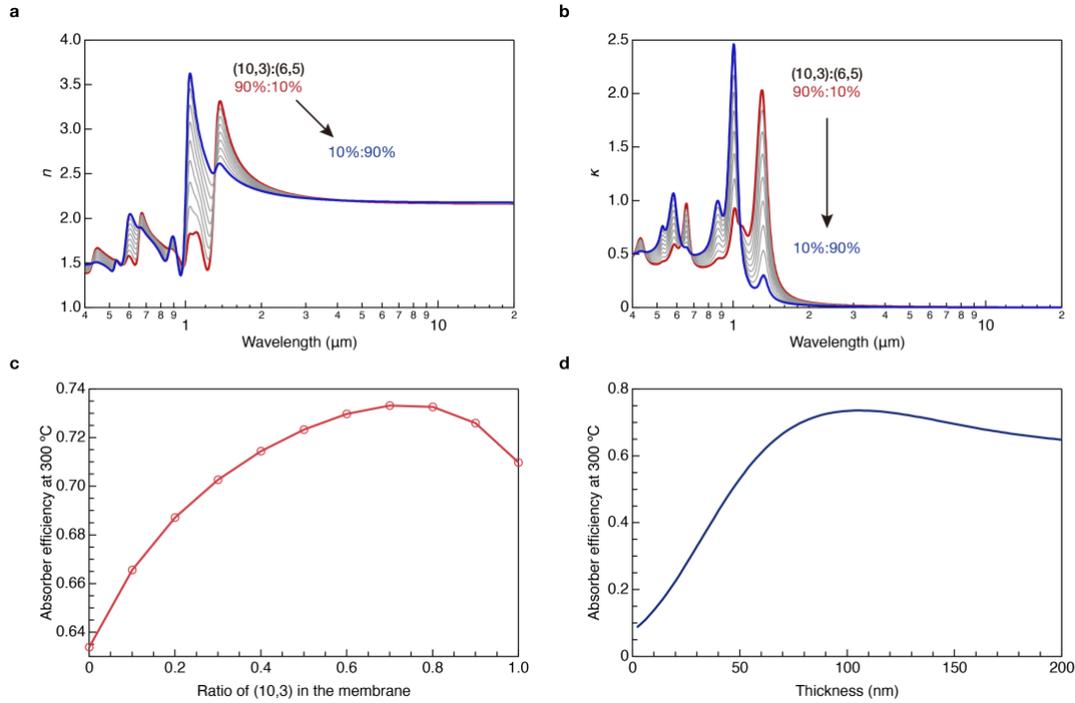

**Figure S6.** (a) Real and (b) imaginary parts of the complex refractive index spectra of the (10,3)–(6,5) mixture membrane with different mixture ratios calculated using the empirical formula. (c) Maximum of absorber efficiency obtained at 300 °C, regarding the mixture ratio. (d) Relation between membrane thickness and absorber efficiency at 300 °C for a (10,3):(6,5) = 70%:30% membrane.

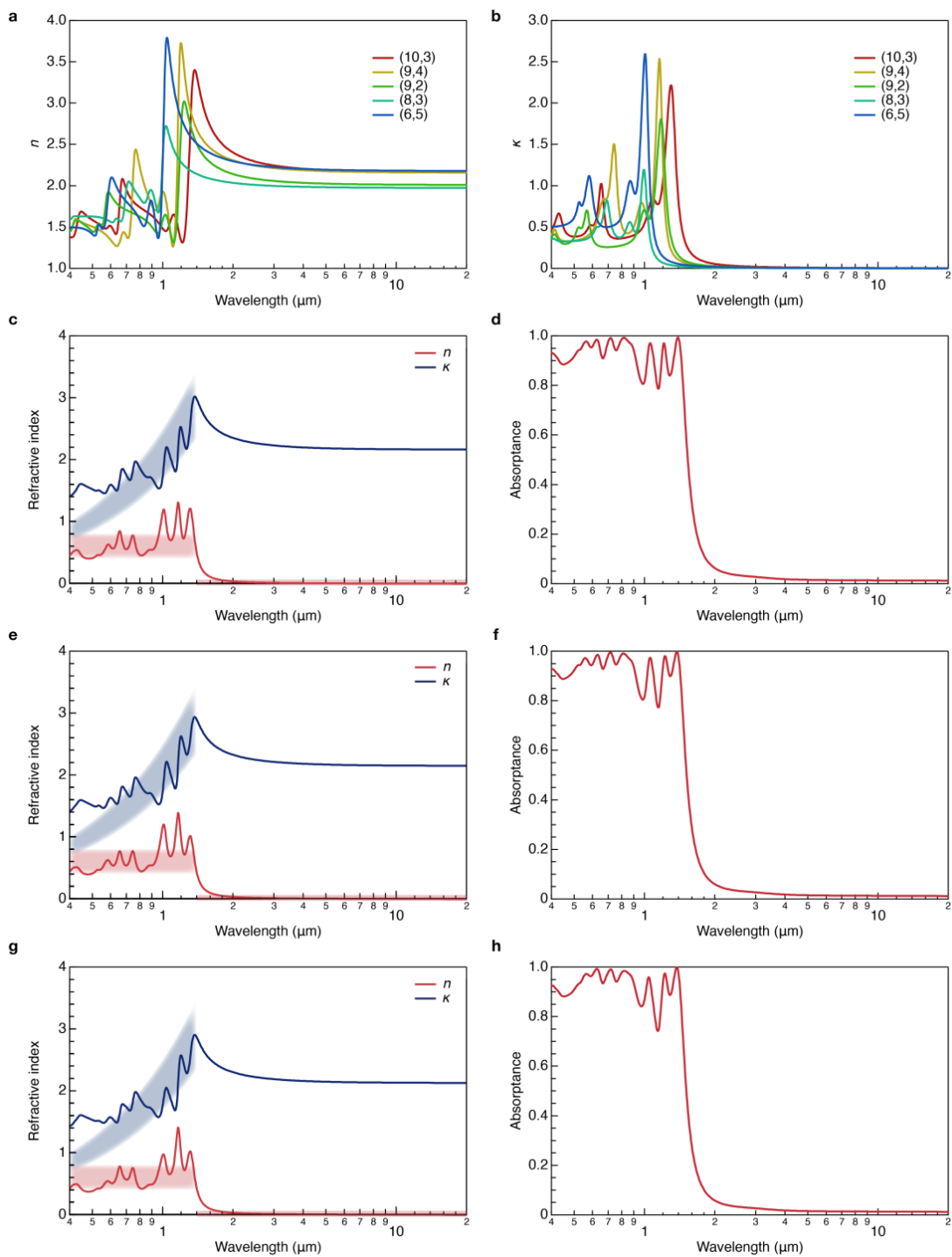

**Figure S7.** (a) Real and (b) imaginary parts of the complex refractive index spectra of the single-chirality membrane calculated using the empirical formula and reported parameters.[63] (c)-(h) Complex refractive index spectra and absorptance spectra of SWCNT membrane comprising of

(c, d) (10,3):(6,5):(9,4) = 50%:20%:30%, (e, f) (10,3):(6,5):(9,4):(9,2) = 40%:20%:30%:10%, (g, h) (10,3):(6,5):(9,4):(9,2):(8,3) = 40%:10%:30%:10%:10%. Blue- and red-filled regions represent the appropriate values of $n$ for layer thickness of 100–150 nm and $\kappa = 2/\pi$ with some tolerance. The thickness is set as 110 nm.

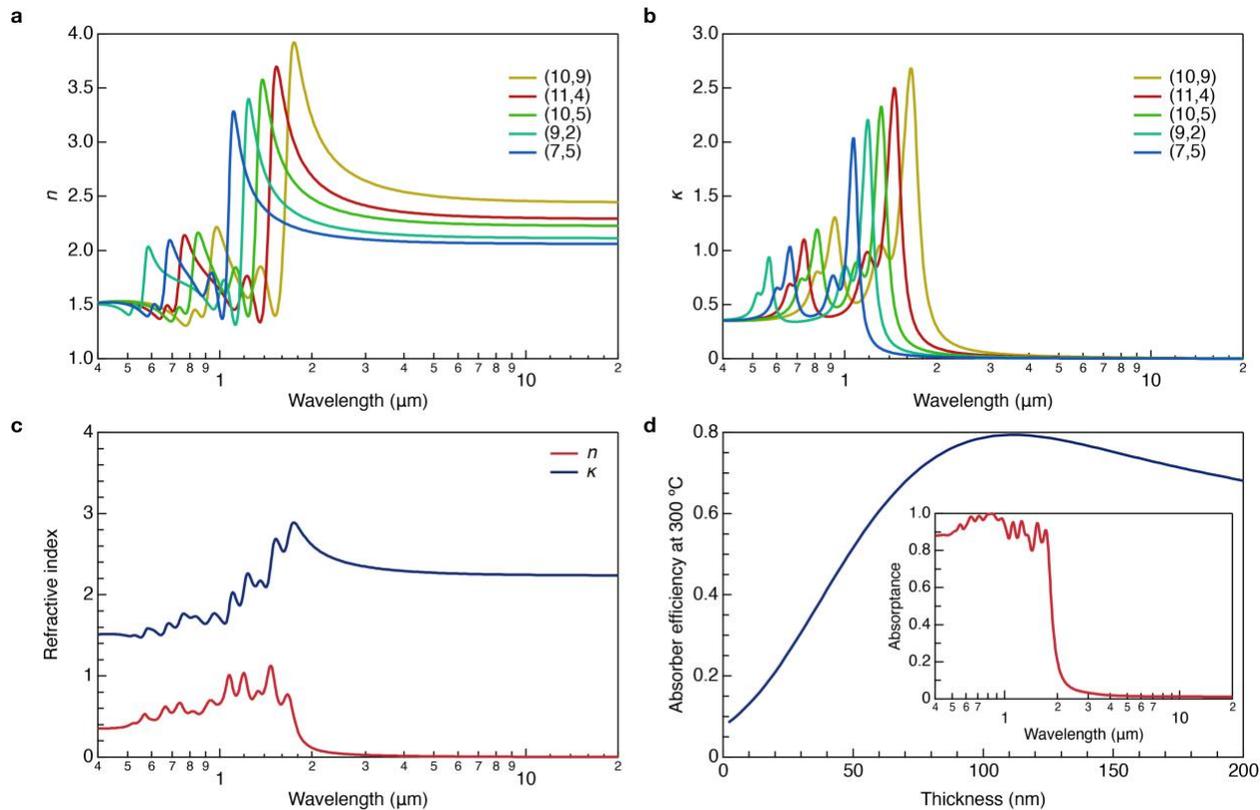

**Figure S8.** (a) Real and (b) imaginary parts of the complex refractive index spectra of the single-chirality membrane calculated using the empirical formula and reported exciton energies.[61,63] (c) Complex refractive index spectra and absorptance spectrum of SWCNT membrane comprising (10,9):(11,4):(10,5):(9,2):(7,5) = 20%:30%:10%:20%:20%. (d) Relation between membrane thickness and absorber efficiency at 300 °C for the optimized membrane. The inset shows the absorptance spectrum when the membrane thickness is 112 nm.

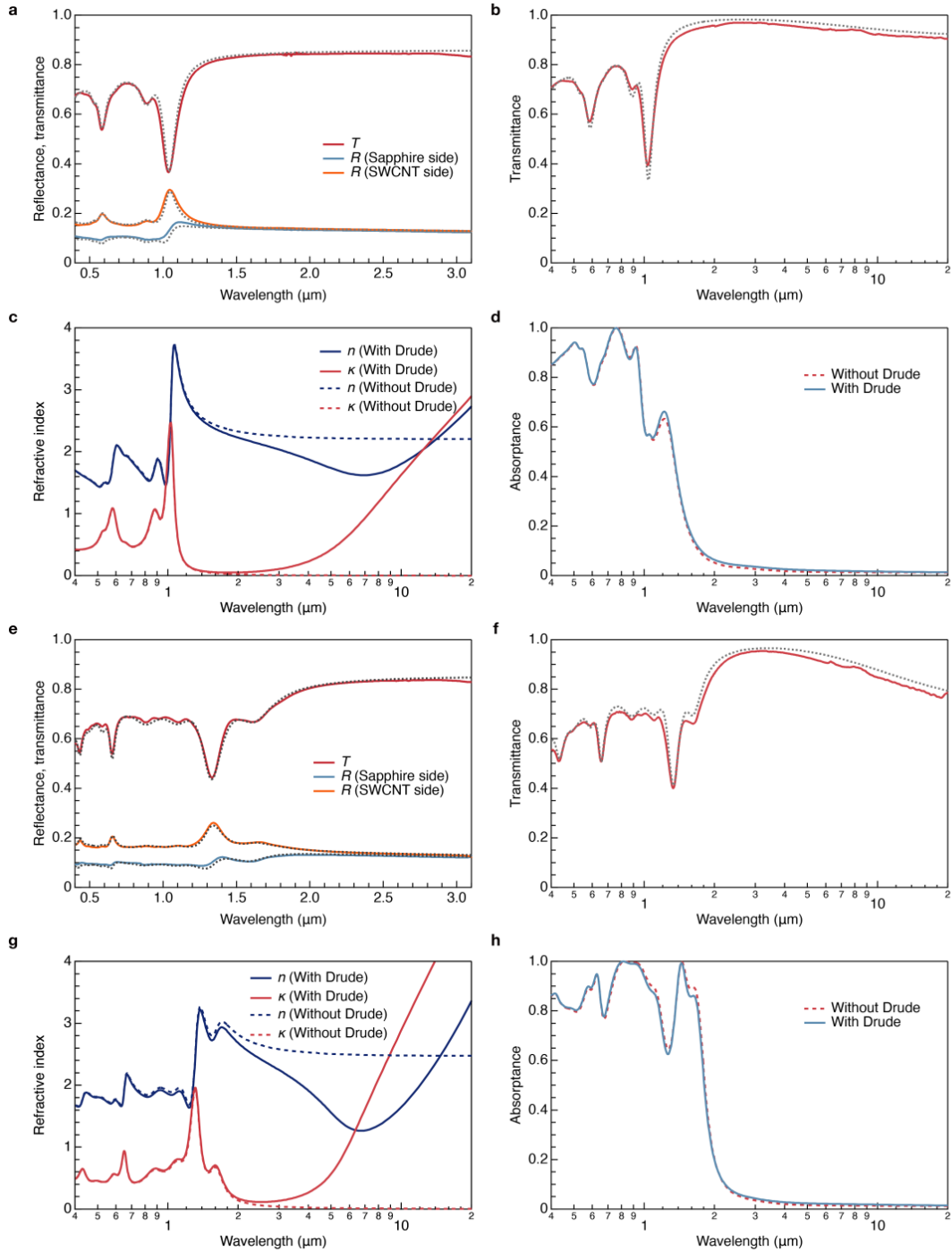

**Figure S9.** (a,b) Reflectance (*R*) and transmittance (*T*) spectra measured for (a) (6,5) on sapphire substrate and (b) freestanding (6,5) membrane. Dotted curves denote the simulation results. (c) Complex refractive spectrum of the (6,5) membrane determined from the measured optical spectra, compared with the ones neglecting the Drude response. (d) Simulated absorptance spectra of (6,5)

membrane on the gold substrate. The thickness is set as 100 nm. (e,f) Reflectance and transmittance spectra measured for (e) (10,3) on sapphire substrate and (f) freestanding (10,3) membrane. Dotted curves denote simulation results. (g) Complex refractive spectrum of the (10,3) membrane determined from the measured optical spectra, compared with the ones neglecting the Drude response. (h) Simulated absorptance spectra of (10,3) membrane on the gold substrate. The thickness is set as 100 nm.

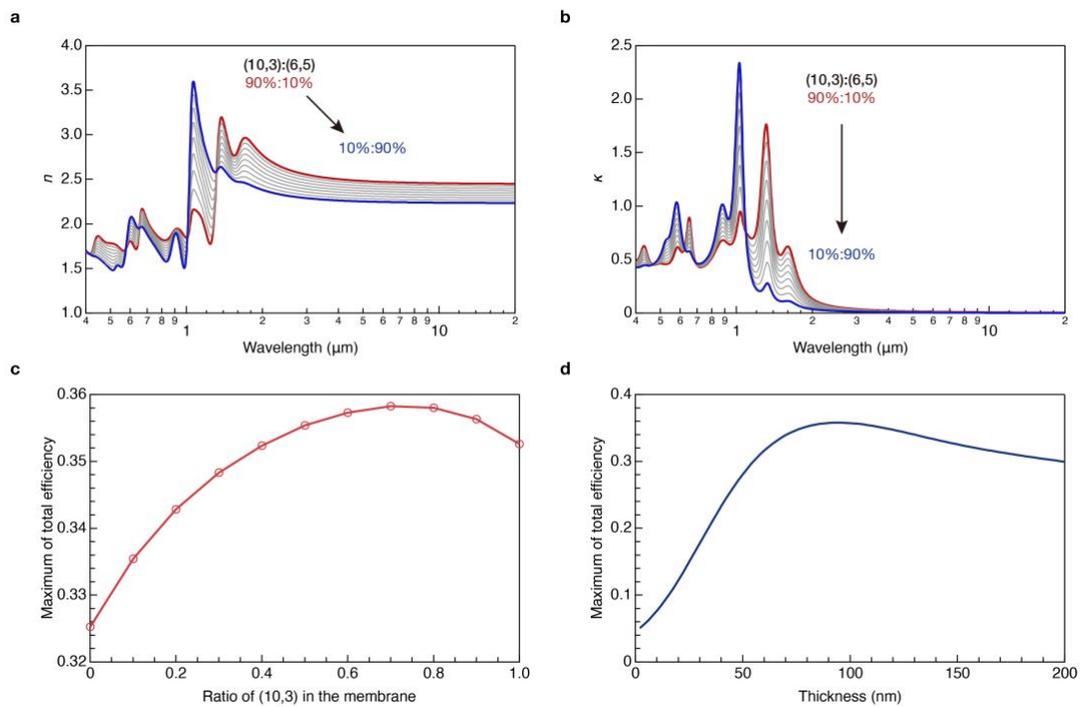

**Figure S10.** (a) Real and (b) imaginary parts of the complex refractive index spectra of the (10,3)–(6,5) mixture membrane with different mixture ratio calculated from the determined one neglecting the Drude response. (c) Maximum of overall conversion efficiency regarding the mixture ratio. (d) Relation between membrane thickness and overall efficiency for a (10,3):(6,5) = 80%:20% membrane.

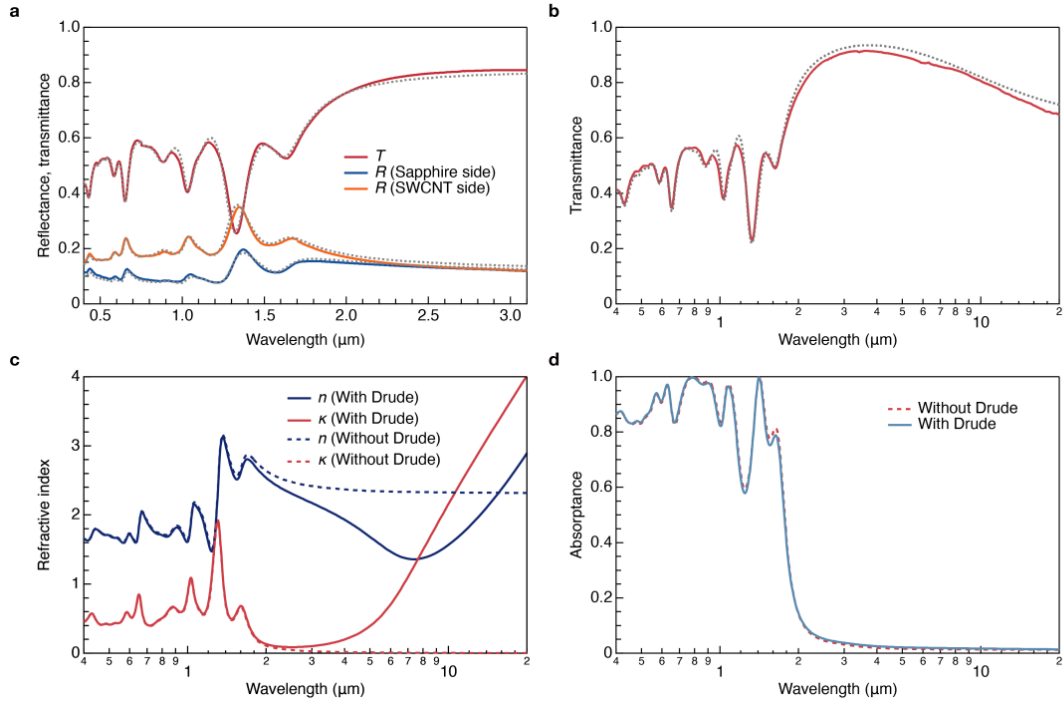

**Figure S11.** (a,b) Reflectance (*R*) and transmittance (*T*) spectra measured for (a) (10,3)–(6,5) on sapphire substrate and (b) freestanding (10,3)–(6,5) membrane. The dotted curves denote simulation results. (c) Complex refractive spectrum of the (10,3)–(6,5) membrane determined from the measured optical spectra, compared with the ones neglecting the Drude response. (d) Simulated absorptance spectra of (10,3)–(6,5) membrane on gold substrate. The thickness is set as 100 nm.

**Supplementary Note 6**

**Upper limit of the equilibrium temperature**

To simulate the theoretical upper limit, we considered a case wherein the absorber received energy input from sunlight and underwent energy exchange only *via* thermal radiation. The upper limit was the temperature at which $\int d\lambda \alpha(\lambda) E_{\text{Solar}}(\lambda) - (\int d\lambda \varepsilon(\lambda) M_{\text{BB}}(\lambda, T_{\text{Abs}}) - \int d\lambda \varepsilon(\lambda) M_{\text{BB}}(\lambda, T_{\text{Envi.}})) = 0$. The first and second terms represent the absorbed solar energy flux and energy fluxes resulting from thermal radiation energy exchange. $T_{\text{Envi.}}$ is the room temperature set as 298 K (25 °C). Figure S12 shows the results with respect to the absorber temperature for a blackbody-absorber and the as-fabricated SWCNT-SSA, respectively. For the blackbody-absorber, the energy flux rapidly drops to zero at approximately 400 K (127 °C), whereas for the SWCNT-SSA, the energy flux drops to zero when the temperature approaches 760 K (487 °C).

In practical experiments, the energy input and output are more complicated. Figure S13a shows the energy input and output of the experiment setup. $Q_{\text{sun}}$ is the absorbed solar energy. $Q_{R1}$, $Q_{R2}$, and $Q_{R3}$ are the energy lost *via* the thermal radiation from the front side of the sample, thermocouple tape, and back side of the sample. $Q_{C1}$ and $Q_{C2}$ are the energy lost *via* the heat conduction to the thermocouple and supporting parts, respectively. Because a gold mirror was placed on the back side of the sample to reflect the back-side radiation, $Q_{R3}$ was neglected. From the energy conservation viewpoint, $|Q_{\text{sun}}| = |Q_{R1} + Q_{R2} + Q_{C1} + Q_{C2}|$. Each term on the right-hand side of the equation increases with temperature. $Q_{R2}$, $Q_{C1}$, and $Q_{C2}$ are the energy losses from the experiment system. By reducing these terms, the measured equilibrium temperature can be closer to the theoretical value and more accurate evaluation can be obtained. For demonstration, we cut the pasted thermocouple to eliminate the $Q_{C1}$ term (Figure S13b), leaving the tape on the surface as the measurement point of the infrared camera. Figure S14 shows the time–temperature curves measured using the improved system. The blackbody-absorber and SWCNT-SSA reached higher temperatures, and the SWCNT-SSA showed a larger temperature increase. The equilibrium temperature measured can be considerably higher and closer to the theoretical value by further reducing the other terms. Large sample area, small relative area of thermocouple tape, and small contact area to the supporting part are essential for achieving these goals.

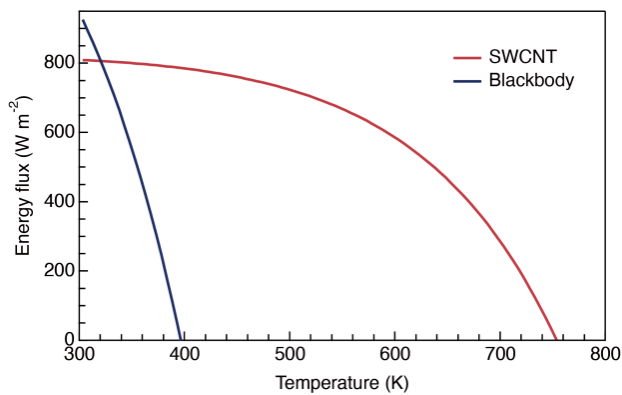

**Figure S12.** Energy flux of the absorbers with respect to the temperature.

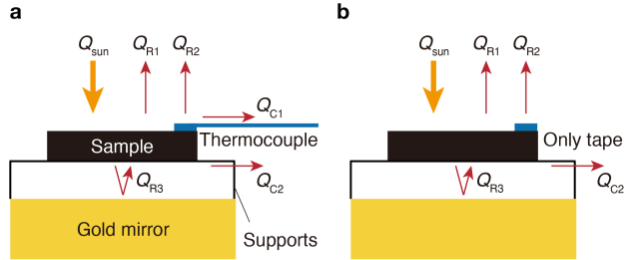

**Figure S13.** (a) Schematic of the energy flow in the measurement system before and (b) after cutting the thermocouple. $Q_{sun}$ is the absorbed solar energy. $Q_{R1}$, $Q_{R2}$, and $Q_{R3}$ are the energy lost *via* the thermal radiation from the front side of the sample, thermocouple tape, and back side of the sample. $Q_{C1}$ and $Q_{C2}$ are the energy lost *via* the heat conduction to the thermocouple and supporting parts, respectively.

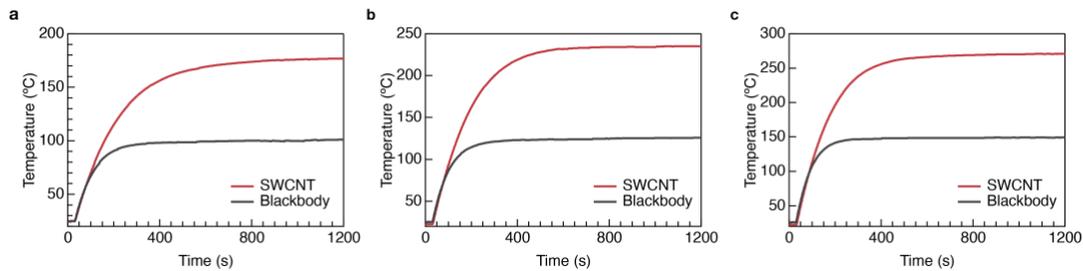

**Figure S14.** (a) Time–temperature curves of blackbody-like absorber and SWCNT-SSA measured using the improved system under 1 SUN (=1 kW m$^{-2}$), (b) 1.5 SUN (=1.5 kW m$^{-2}$) and (c) 2 SUN (=2 kW m$^{-2}$) illumination.

**Supplementary Note 7**

**Multiple reflection model for angle-dependent optical simulation**

For s-polarized light, the reflectance spectrum was calculated as follows:[77]

$$R^s = \left| \frac{\tilde{r}^s_{AC} + \tilde{r}^s_{CS} \exp(-i2\tilde{\beta}^s)}{1 + \tilde{r}^s_{AC}\tilde{r}^s_{CS} \exp(-i2\tilde{\beta}^s)} \right|^2, \quad (S8)$$

where $\tilde{r}^s_{AC}$ and $\tilde{r}^s_{CS}$ are the amplitude reflection coefficients of s-polarized light at the air–SWCNT and SWCNT–substrate interfaces, respectively. $\tilde{\beta}^s$ is the absorption-induced phase variation of s-polarized light during propagation in SWCNT membranes. These values can be expressed as follows:

$$\tilde{r}^s_{AC} = \frac{\cos\theta - (\tilde{n}^{*\,2}_{in} - \sin^2\theta)^{1/2}}{\cos\theta + (\tilde{n}^{*\,2}_{in} - \sin^2\theta)^{1/2}}, \quad (S9)$$

$$\tilde{r}^s_{CS} = \frac{(\tilde{n}^{*\,2}_{in} - \sin^2\theta)^{1/2} - (\tilde{n}^2_{sub} - \sin^2\theta)^{1/2}}{(\tilde{n}^{*\,2}_{in} - \sin^2\theta)^{1/2} + (\tilde{n}^2_{sub} - \sin^2\theta)^{1/2}}, \quad (S10)$$

$$\tilde{\beta}^s = \frac{2\pi d}{\lambda}(\tilde{n}^{*\,2}_{in} - \sin^2\theta)^{1/2}, \quad (S11)$$

where $\tilde{n}_{in}$ is the complex refractive index of the SWCNT membrane to the in-plane polarized light and $\tilde{n}^*_{in}$ is its complex conjugate. $\lambda$ is the wavelength, $d$ is the thickness of the SWCNT membrane, and $\tilde{n}_{sub}$ is the complex refractive index of the substrate. For p-polarized light, $\tilde{r}^s_{AC}$, $\tilde{r}^s_{CS}$, and $\tilde{\beta}^s$ in Eq. (S8) are replaced with $\tilde{r}^p_{AC}$, $\tilde{r}^p_{CS}$ and $\tilde{\beta}^p$, respectively. These terms are given by:

$$\tilde{r}^p_{AC} = \frac{\tilde{n}^*_{in}\tilde{n}^*_{out}\cos\theta - (\tilde{n}^{*\,2}_{out} - \sin^2\theta)^{1/2}}{\tilde{n}^*_{in}\tilde{n}^*_{out}\cos\theta + (\tilde{n}^{*\,2}_{out} - \sin^2\theta)^{1/2}}, \quad (S12)$$

$$\tilde{r}^p_{CS} = \frac{\tilde{n}^2_S(\tilde{n}^{*\,2}_{out} - \sin^2\theta)^{1/2} - \tilde{n}^*_{in}\tilde{n}^*_{out}(\tilde{n}^2_{sub} - \sin^2\theta)^{1/2}}{\tilde{n}^2_S(\tilde{n}^{*\,2}_{out} - \sin^2\theta)^{1/2} + \tilde{n}^*_{in}\tilde{n}^*_{out}(\tilde{n}^2_{sub} - \sin^2\theta)^{1/2}}, \quad (S13)$$

$$\tilde{\beta}^p = \frac{2\pi d}{\lambda}\left(\frac{\tilde{n}_{in}^*}{\tilde{n}_{out}^*}\right)\left(\tilde{n}_{out}^{*2} - \sin^2\theta\right)^{1/2}, \tag{S14}$$

Where $\tilde{n}_{out}$ is the complex refractive index of the SWCNT membrane to light polarized perpendicularly to the membrane surface and $\tilde{n}_{out}^*$ is its complex conjugate. The reflectance of the unpolarized light is averaged from those of the *s*- and *p*-polarized lights.

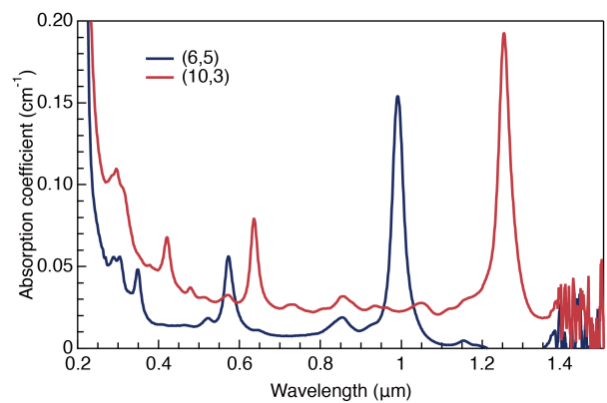

**Figure S15.** Typical absorption spectra of (6,5) SWCNT and (10,3) SWCNT dispersions.